\documentclass[aps,prd,preprintnumbers,nofootinbib,superscriptaddress,preprint]{revtex4-1}
\usepackage{graphicx}
\usepackage{amssymb}
\usepackage{amsmath}
\usepackage{color}
\usepackage{booktabs}
\usepackage{dcolumn}

\graphicspath{{fig/}}

\newcommand{\beq}{\begin{eqnarray}}
\newcommand{\eeq}{\end{eqnarray}}

\def\fsl#1{\setbox0=\hbox{$#1$}
   \dimen0=\wd0
   \setbox1=\hbox{/} \dimen1=\wd1
   \ifdim\dimen0>\dimen1
      \rlap{\hbox to \dimen0{\hfil/\hfil}}
      #1
   \else
      \rlap{\hbox to \dimen1{\hfil$#1$\hfil}}
      /
   \fi}

\begin{document}

\preprint{UTHEP-743, UTCCS-P-128, HUPD-1916}

\title{$K_{l3}$ form factors at the physical point on (10.9 fm)$^3$ volume}

\date{ \today
}


\author{Junpei~Kakazu}
\affiliation{Faculty of Pure and Applied Sciences, University of Tsukuba, \\ Tsukuba, Ibaraki 305-8571, Japan}

\author{Ken-ichi~Ishikawa}
\affiliation{Graduate School of Advanced Science and Engineering, Hiroshima University, Higashi-Hiroshima, Hiroshima 739-8526, Japan}

\author{Naruhito~Ishizuka}
\affiliation{Center for Computational Sciences, University of Tsukuba, \\ Tsukuba, Ibaraki 305-8577, Japan}

\author{Yoshinobu~Kuramashi}
\affiliation{Center for Computational Sciences, University of Tsukuba, \\ Tsukuba, Ibaraki 305-8577, Japan}

\author{Yoshifumi~Nakamura}
\affiliation{RIKEN Center for Computational Science, Kobe, Hyogo 650-0047, Japan}

\author{Yusuke~Namekawa}
\affiliation{Institute of Particle and Nuclear Studies, High Energy Accelerator Research Organization (KEK), Tsukuba 305-0801, Japan}

\author{Yusuke~Taniguchi}
\affiliation{Center for Computational Sciences, University of Tsukuba, \\ Tsukuba, Ibaraki 305-8577, Japan}

\author{Naoya~Ukita}
\affiliation{Center for Computational Sciences, University of Tsukuba, \\ Tsukuba, Ibaraki 305-8577, Japan}

\author{Takeshi~Yamazaki}
\affiliation{Faculty of Pure and Applied Sciences, University of Tsukuba, \\ Tsukuba, Ibaraki 305-8571, Japan}
\affiliation{Center for Computational Sciences, University of Tsukuba, \\ Tsukuba, Ibaraki 305-8577, Japan}

\author{Tomoteru~Yoshi\'e}
\affiliation{Center for Computational Sciences, University of Tsukuba, \\ Tsukuba, Ibaraki 305-8577, Japan}

\collaboration{PACS Collaboration}

\begin{abstract}

We present the calculation of the $K_{l3}$ form factors 
with $N_f = 2 + 1$ nonperturbatively $O(a)$-improved Wilson quark action
and Iwasaki gauge action at the physical point
on a large volume of (10.9 fm)$^3$ at one lattice spacing of
$a = 0.085$ fm.
We extract the form factors from 3-point functions 
with three different time separations between the source and sink operators
to confirm suppression of excited state contributions.
The form factors are calculated in very close to the zero
momentum transfer, $q^2 = 0$, thanks to the large volume, so that
stable interpolations to $q^2 = 0$ are carried out.
Using our form factors, 
we obtain the form factor at $q^2 = 0$, 
$f_+(0) = 0.9603(16)(^{+14}_{\ -4})(44)(19)(1)$,
where the first, second, and fifth errors are
statistical, systematic errors from fit functions
and the isospin breaking effect, respectively.
The third and fourth errors denote the finite lattice spacing effects 
estimated from the renormalization factor and contribution
beyond the leading order SU(3) chiral perturbation theory (ChPT).
The result of $f_+(0)$ yields the Cabibbo-Kobayashi-Maskawa (CKM) matrix 
element, $|V_{us}| = 0.2255(13)(4)$, where the first error comes from 
our calculation and the second from the experiment.
This value is consistent with the ones determined from the unitarity
of the CKM matrix and the $K_{l2}$ decay within one standard deviation,
while it is slightly larger than recent lattice calculations
by at most 1.5 $\sigma$.
Furthermore, we evaluate the shape of the form factors and
the phase space integral from our results.
We confirm that those results are consistent with the experiment,
and also $|V_{us}|$ determined with our phase space integral
agrees with the one in the above.

\end{abstract}

\maketitle

\section{Introduction}

Search for signals beyond the standard model (BSM) is an important
task in the field of the particle physics.
In indirect search for the BSM physics, it is necessary to precisely compare 
physical quantities obtained from 
experiments and their predictions in the standard model (SM).
Currently one of the indirect searches is carried out through the 
CKM matrix element $|V_{us}|$.
Its SM prediction is evaluated by
the unitarity of the CKM matrix in the first row, {\it i.e.},
$|V_{ud}|^2 + |V_{us}|^2 + |V_{ub}|^2 = 1$.
Using the precisely determined value of 
$|V_{ud}| = 0.97420(21)$~\cite{Marciano:2005ec,Tanabashi:2018oca},
the SM prediction is $|V_{us}| = 0.2257(9)$, 
where $|V_{ub}|$ is neglected in the estimate 
due to $|V_{ub}| \ll |V_{ud}|$.

Experimentally, $|V_{us}|$ is related to kaon decay processes, such as
the $K_{l2}$ decay, $K \to l \nu$,
and the $K_{l3}$ decay, $K \to \pi l \nu$, processes,
where $K$, $\pi$, $l$, and $\nu$ are the kaon, pion, lepton 
and neutrino, respectively.
In both cases, $|V_{us}|$ is not determined from the experiments only,
and lattice QCD calculation also plays an important role,
which is the first principle calculation of the strong interaction.
For the $K_{l2}$ decay, the ratio of the decay constants for
the kaon and pion, $F_K/F_\pi$, is required to determine 
the value of $|V_{us}|$.
Using current lattice results,
for example $F_K/F_\pi = 1.1933(29)$ in Ref.~\cite{Tanabashi:2018oca},
$|V_{us}|$ from the $K_{l2}$ decay well agrees with the SM
prediction in the above.

For the $K_{l3}$ decay, $|V_{us}|$ is related to 
the decay rate of the $K$ decay $\Gamma_{K_{l3}}$ as,
\begin{equation}
\Gamma_{K_{l3}} = C_{K_{l3}} ( |V_{us}| f_+(0) )^2 I^l_K,
\label{eq:decay_width_kl3}
\end{equation}
where $f_+(0)$ is the value of the $K_{l3}$ form factor at $q^2 = 0$
with $q$ being the momentum transfer,
$C_{K_{l3}}$ is a known factor including the electromagnetic 
correction and the SU(2) breaking effect,
and $I^l_K$ is the phase space integral calculated from the shape of
the $K_{l3}$ form factors.
The experiment determines $\Gamma_{K_{l3}}$ and also $I^l_K$.
The value of $f_+(0)$, however, is not obtained from the experiment.
Currently a precise calculation of $f_+(0)$ can be performed by lattice QCD.

So far various lattice QCD calculations with dynamical quarks 
have been carried out to evaluate the value of $f_+(0)$~\cite{Bazavov:2012cd,Boyle:2007qe,Boyle:2013gsa,Boyle:2015hfa,Aoki:2017spo,Bazavov:2013maa,Carrasco:2016kpy,Bazavov:2018kjg,Dawson:2006qc,Lubicz:2009ht}.
The most recent study in the $N_f = 2 + 1 + 1$ QCD~\cite{Bazavov:2018kjg} using 
the staggered-type quark action
reported that using their value of $f_+(0)$ there is a clear deviation 
of $|V_{us}|$ in more than 2 $\sigma$ from the ones in
the SM prediction and the $K_{l2}$ decay.
Another $N_f = 2 + 1 + 1$ calculation using the twisted quark action~\cite{Carrasco:2016kpy} obtained a similar result.
Therefore, it is an urgent task for the search for BSM signals 
to confirm those results by several lattice calculations 
using different types of quark actions with 
small statistical and systematic uncertainties.

For this purpose we calculate the $K_{l3}$ form factors in the $N_f = 2 + 1$ QCD
using a nonperturbatively $O(a)$-improved Wilson quark action and Iwasaki gauge action
at one lattice spacing of $a = 0.085$ fm.
The gauge configurations employed in this work are a subset of 
the PACS10 configurations~\cite{Ishikawa:2018jee}.
Our calculation is carried out at the physical light and strange
quark masses, and on a larger physical volume of (10.9 fm)$^3$ than
typical current lattice QCD calculations.
Thus, our result significantly suppresses the uncertainties coming from
the chiral extrapolation and finite volume effect.
Another advantage using the large volume is
that it is possible to access the small $q^2$ region without resort to 
the twisted boundary condition.
Thanks to the large volume, one piece of our data is very close to $q^2 = 0$,
so that we can perform reliable interpolations of the form factors to $q^2 = 0$.
Using our result of $f_+(0)$, we determine $|V_{us}|$ and compare it with
the SM prediction, the $K_{l2}$ decay, and also the previous
lattice QCD results.
Furthermore, since we calculate the form factors in a wide range of $q^2$,
the shape of the form factors and also the phase space
integral are successfully evaluated from our results.
Those values are compared with the experiment and previous lattice QCD results.
We also determine $|V_{us}|$ using our result of the phase space integral.
Our preliminary result has been already reported in 
Ref.~\cite{Kakazu:2019hpl}.

This paper is organized as follows.
Section~\ref{sec:method} explains our calculation method of the form factors
from meson 2- and 3-point functions.
In Sec.~\ref{sec:set_up} simulation parameters and technical 
details of our calculation
are presented.
The result of the form factors is shown in Sec.~\ref{sec:form_factors}.
The interpolations of the form factors are discussed in Sec.~\ref{sec:interpolation}.
Our results for $f_+(0)$, and the shape of the form factors,
and phase space integral are also presented in this section.
The result of $|V_{us}|$ and its comparison with other determinations
are discussed in Sec.~\ref{sec:v_us}.
Section~\ref{sec:conclusions} is devoted to conclusion.
Appendices explain interpolating functions based on the SU(3) ChPT
used in our analysis and tables for some interpolation results.

\section{Calculation method}
\label{sec:method}

The $K_{l3}$ form factors $f_+(q^2)$ and $f_-(q^2)$ are defined by the matrix element of the weak vector current $V_{\mu}$ as,
\begin{eqnarray}
\langle \pi (\vec{p}_{\pi}) \left | V_{\mu} \right | K(\vec{p}_{K}) \rangle = ({p}_{K}+{p}_{\pi})_{\mu}f_{+}(q^2)+ ({p}_{K}-{p}_{\pi})_{\mu}f_{-}(q^2),
\label{eq:def_matrix_element}
\end{eqnarray}
where $q=p_{K}-p_{\pi}$ is the momentum transfer.
The scalar form factor $f_0(q^2)$ is defined by $f_+(q^2)$ and $f_-(q^2)$
as,
\begin{eqnarray}
f_{0}(q^2) =f_{+}(q^2) + \frac{-q^2}{{m^2_{K}}-{m^2_{\pi}}}f_{-}(q^2)
= f_{+}(q^2)\left(1+ \frac{-q^2}{{m^2_{K}}-{m^2_{\pi}}}\xi(q^2)\right), 
\label{eq:f0}
\end{eqnarray}
where $\xi(q^2)=f_{-}(q^2)/f_{+}(q^2)$,
and $m_\pi$ and $m_K$ are the masses for $\pi$ and $K$, respectively.
At $q^2=0$, the two form factors $f_+(q^2)$ and $f_0(q^2)$ 
give the same value, $f_+(0)=f_0(0)$.

The $K_{l3}$ form factors are calculated from
3-point function $C_{\mu}^{K\pi}(\vec{p} ,t)$
with the weak vector current given by
\begin{eqnarray}
C_{\mu}^{K\pi}(\vec{p }, t)&=&
\langle 0 | O_K( \vec{0},t_f) V_\mu(\vec{p},t) O^\dagger_\pi( \vec{p},t_i)| 0 \rangle,
\label{eq:def_3-pt}
\end{eqnarray}
where 
\begin{eqnarray}
O_\pi(\vec{p},t) &=& \sum_{\vec{x}}
\overline{u}(\vec{x},t)\gamma_5 d(\vec{x},t) e^{i\vec{p}\cdot\vec{x}} ,\\
O_K(\vec{p},t) &=& \sum_{\vec{x}}
\overline{s}(\vec{x},t)\gamma_5 d(\vec{x},t) e^{i\vec{p}\cdot\vec{x}} ,\\
V_\mu(\vec{p},t) &=& \sum_{\vec{x}}
\overline{u}(\vec{x},t)\gamma_\mu s(\vec{x},t) e^{i\vec{p}\cdot\vec{x}} .
\end{eqnarray}
We use only the periodic boundary condition in the spatial directions
for quark propagators in contrast to 
the recent calculations of the $K_{l3}$ form factors
using the twisted boundary condition~\cite{Bazavov:2018kjg,Aoki:2017spo,Carrasco:2016kpy},
because the spatial extent $L$ in our calculation is large enough to 
obtain the form factors near the $q^2=0$ region.
Thus, $\vec{p}$ is labeled by an integer vector $\vec{n}_p$ with $p=\vert{\vec p}\vert$ as 
$\vec{p} = (2\pi/L) \vec{n}_p$.
While we have also calculated the 3-point functions with moving $K$ and 
$\pi$ at rest, their form factors
are much noisier than the ones from $C_\mu^{K\pi}(\vec{p},t)$.
Therefore, we will not discuss those results in this paper.

For the renormalization factor of the vector current $Z_V$,
we compute 3-point functions for the $\pi$ and $K$
electromagnetic form factors at $q^2 = 0$ in a similar way, which are given by
\begin{eqnarray}
C^{\pi\pi}_4(t) &=& \langle 0 | O_\pi(\vec{0},t_f)V^{\rm em}_4(t)O^\dagger_\pi(\vec{0},t_i)|0\rangle ,
\label{eq:3-pt_em_pi}\\
C^{KK}_4(t) &=& \langle 0 | O_K(\vec{0},t_f)V^{\rm em}_4(t)O^\dagger_K(\vec{0},t_i)|0\rangle ,
\label{eq:3-pt_em_K}
\end{eqnarray}
where $V_4^{\rm em}(t)$ is the temporal component of the electromagnetic 
current.

The 2-point functions for $\pi$ and $K$ are calculated as,
\begin{eqnarray}
C^{\pi}(\vec{p},t-t_i)&=&
\langle 0 | O_\pi( \vec{p},t) O^\dagger_\pi( \vec{p},t_i)| 0 \rangle,
\label{eq:def_2-pt_pi}\\
C^{K}(\vec{p},t-t_i)&=&
\langle 0 | O_K( \vec{p},t) O^\dagger_K( \vec{p},t_i)| 0 \rangle .
\label{eq:def_2-pt_K}
\end{eqnarray}
We average the 2-point functions with the periodic and 
anti-periodic boundary conditions in the temporal direction
to make the periodicity in the temporal direction 
of the 2-point functions effectively doubled.
The asymptotic form of $C^{X}(\vec{p},t)$ for $X = \pi$ and $K$
in the $t \gg 1$ region is given by
\begin{eqnarray}
C^{X}(\vec{p},t)= \frac{Z_X^2}{2E_X(p)} ( e^{-E_X(p) t}+e^{-E_X(p) (2T - t)} ),
\label{eq:asym_form_two-pt}
\end{eqnarray}
with $E_X(p) = \sqrt{m_X^2 + p^2}$ and the temporal extent $T$.
The mass $m_X$ and amplitude $Z_X$ are obtained from a fit 
of $C^{X}(\vec{0},t)$ in a large $t$ region with the
asymptotic form.

The matrix element in Eq.~(\ref{eq:def_matrix_element})
is obtained from the ground state contribution of 
$C_{\mu}^{K\pi}(\vec{p }, t)$, which needs to avoid
excited state contributions by investigating 
time dependences of $C_{\mu}^{K\pi}(\vec{p},t)$.
To do this, we define a ratio $R_\mu^{\rm BC}(\vec{p},t)$,
which has the following time dependence as,
\begin{eqnarray}
R^{\rm BC}_\mu(\vec{p},t) &=& \frac{N_\mu (\vec{p}) C_{\mu, {\rm BC}}^{K\pi}(\vec{p},t)}
{C^\pi(\vec{p},t-t_i)C^K(\vec{0},t_f-t)}
\label{eq:def_Rmu}
\\
&=&
\frac{N_\mu (\vec{p})}{Z_V Z_\pi Z_K}
\left( \langle \pi (\vec{p}) \left | V_{\mu} \right | K(\vec{0}) \rangle   
 + \Delta A_\mu(\vec{p},t) + b_{\rm BC} \Delta B_\mu(\vec{p},t) +\cdots \right),
\label{eq:Rmu_time_dep}
\end{eqnarray}
where $N_4(\vec{p}) = 1$ and $N_i(\vec{p}) = 1/p_i$ with $i = 1,2,3$,
and $Z_\pi$ and $Z_K$ are defined in Eq.~(\ref{eq:asym_form_two-pt}).
$C_{\mu, {\rm BC}}^{K\pi}(\vec{p},t)$ is 
the 3-point function in Eq.~(\ref{eq:def_3-pt}) with
the (anti-)periodic boundary condition in the temporal direction,
which is represented by ${\rm BC = (A)PBC}$ in the following.
In Eq.~(\ref{eq:Rmu_time_dep}) it is assumed that $t_i \le t \le t_f$
and two excited state contributions for
the radial excited mesons and wrapping around effect, expressed by
$\Delta A_\mu(\vec{p},t)$ and $\Delta B_\mu(\vec{p},t)$, respectively, are
leading contributions of excited states in the ratio.
Other excited state contributions are denoted by
the dots $(\cdots)$ term.
The sign of the wrapping around effect in the temporal direction depends on 
the temporal boundary condition of 
$C_{\mu, {\rm BC}}^{K\pi}(\vec{p}, t)$,
{\it i.e.}, $b_{\rm PBC} = 1$ and $b_{\rm APBC} = -1$,
because in the wrapping around contribution one of the mesons 
in $C_{\mu, {\rm BC}}^{K\pi}(\vec{p},t)$ crosses the temporal boundary.
A similar wrapping around effect was discussed in the 3-point function
of the $B_K$ calculation~\cite{Aoki:2008ss}.

The time dependence of the two excited state contributions is given by
\begin{eqnarray}
\Delta A_\mu(\vec{p},t) &=& 
A_\mu^\pi(\vec{p}) e^{-(E_\pi^\prime(p) - E_\pi(p))(t-t_i)} +
A_\mu^K(\vec{p}) e^{-(m_K^\prime - m_K)(t_f-t)} ,
\label{eq:def_radial_excited_effect}
\\
\Delta B_\mu(\vec{p},t) &=&
B_\mu^\pi(\vec{p}) e^{-E_\pi(p)(T+2t_i-2t)} +
B_\mu^K(\vec{p}) e^{-m_K(T+2t-2t_f)} ,
\label{eq:def_wrappind_around_effect}
\end{eqnarray}
where $E_\pi^\prime(p) = \sqrt{(m_\pi^\prime)^2 + p^2}$, and
$m_X^\prime$ is the mass of the radial excitation of $X = \pi$ and $K$.
In the second equation, we assume that the finite volume effect in 
the energy of the $\pi K$ scattering state is negligible in our volume. 
In a small $p$, the first term of the right hand side in 
Eq.~(\ref{eq:def_wrappind_around_effect}) has a non-negligible effect
in $R_\mu^{\rm BC}(\vec{p},t)$, which will be presented later.
This is because $m_\pi T = 7.5$ is not enough to suppress 
the wrapping around effect at the physical $m_\pi$.
We remove the wrapping around effect
$\Delta B_\mu(\vec{p},t)$ by averaging the ratios
$R_\mu^{\rm PBC}(\vec{p},t)$ and $R_\mu^{\rm APBC}(\vec{p},t)$.
On the other hand, another excited state contrition 
$\Delta A_\mu(\vec{p},t)$ remains in the averaged ratio.
This contribution needs to be removed, and 
it will be discussed in a later section.

\section{Set up}
\label{sec:set_up}

We use the configurations generated with the Iwasaki 
gauge action~\cite{Iwasaki:2011jk}
and the stout-smeared Clover quark action at the physical point
on $(L/a)^3\times T/a = 128^3\times 128$ lattice corresponding to (10.9 fm)$^4$.
These configurations are a subset of the PACS10 configurations.
Parameters for the gauge configuration generation are found
in Ref.~\cite{Ishikawa:2018jee}.
The bare coupling $\beta = 1.82$ corresponds to
$a^{-1} = 2.3162(44)$ GeV~\cite{Ishikawa:2019qwn} 
determined from the $\Xi$ baryon mass input.
The hopping parameters for the light and strange quarks are
$(\kappa_l, \kappa_s) = (0.126117, 0.124902)$,
and the coefficient of the clover term is $c_{\rm SW} = 1.11$,
which is nonperturbatively determined in the Schr\"odinger functional
(SF) scheme~\cite{Taniguchi:2012kk}.
It is employed the six-stout-smeared link~\cite{Morningstar:2003gk} 
with $\rho = 0.1$ in the quark actions. 
We use the same quark actions for the measurement of the $K_{l3}$
form factors.
The measured $\pi$ and $K$ masses, 
$m_\pi = 0.13511(72)$ GeV and $m_K = 0.49709(35)$ GeV,
in this calculation are consistent with the ones in our spectrum 
paper~\cite{Ishikawa:2018jee}.

The measurements for the 2-point and 3-point functions 
are performed using 20 configurations separated by 
10 molecular dynamics trajectories.
To reduce the calculation cost of the measurements,
the $Z(2)\otimes Z(2)$ random source~\cite{Boyle:2008yd} at 
the source time slice $t_i$ is employed,
where random numbers are spread in the color and spin spaces
as well as the spatial volume.
For example, 
the operator $O_\pi(\vec{p},t)$ at the source time slice $t_i$
in Eq.~(\ref{eq:def_3-pt}) is replaced by
\begin{equation}
O_\pi(\vec{p},t,\eta) = \frac{1}{N_r}\sum_j
\left[ \sum_{\vec{x}}\overline{u}(\vec{x},t)\eta_j^\dagger(\vec{x})e^{i\vec{p}\cdot\vec{x}} \right]
\gamma_5
\left[ \sum_{\vec{y}}d(\vec{y},t)\eta_j(\vec{y}) \right] ,
\label{eq:random_source}
\end{equation}
where $N_r$ is the number of the random source, and
the color and spin indices are omitted.
The $Z(2)\otimes Z(2)$ random source $\eta_j(\vec{x})$ 
satisfies the following condition as,
\begin{equation}
\frac{1}{N_r}\sum_j \eta_j^\dagger(\vec{x})\eta_j(\vec{y}) \xrightarrow[N_r \to \infty]{} \delta(\vec{x}-\vec{y}) .
\end{equation}

We use the sequential source method at the sink time slice $t_f$
in $C^{K\pi}_{\mu, {\rm BC}}(\vec{p},t)$.
The quark propagators are calculated with the periodic boundary condition
in the spatial and also temporal directions
in $C^{K\pi}_{\mu, {\rm PBC}}(\vec{p},t)$.
On the other hand, in $C^{K\pi}_{\mu, {\rm APBC}}(\vec{p},t)$,
though the spatial boundary condition is periodic for all the quark propagators,
one of the three quark propagators needs to 
be calculated with the anti-periodic boundary condition
in the temporal direction.
In this work we choose the quark propagator which connects 
the source operator with the sink one.
This choice is suitable for our purpose to remove the wrapping around effect
$\Delta B_\mu(\vec{p},t)$ in Eq.~(\ref{eq:Rmu_time_dep}), 
because in this case the effect has a desirable boundary condition 
dependence as in Eq.~(\ref{eq:Rmu_time_dep}).
A similar technique using combination of quark propagators with 
the periodic and anti-periodic boundary conditions in the temporal direction 
was employed in the $B_K$ calculation to effectively double 
the periodicity of the 3-point function~\cite{Arthur:2012yc,Blum:2014tka}.
It is noted that partially quenched effects due to
the different boundary condition from the sea quarks are expected to
be exponentially suppressed as in the twisted boundary condition
discussed in Ref.~\cite{Sachrajda:2004mi}.

In each momentum, the quark propagator of the random momentum source
corresponding to the first square brackets in Eq.~(\ref{eq:random_source}) 
is calculated.
To improve the statistical error of $C^{K\pi}_{\mu,{\rm BC}}(\vec{p},t)$ 
in a finite momentum, we average $C^{K\pi}_{\mu,{\rm BC}}(\vec{p},t)$
in each $n_p = (Lp/2\pi)^2$ with several momentum assignments.
The number of the momentum assignment is listed in Table~\ref{tab:q2}
together with the values of $q^2 = -(m_K-E_\pi(p))^2+p^2$
calculated using the measured $m_K$ and $m_\pi$
with $E_\pi(p) = \sqrt{m_\pi^2 + p^2}$.
The value of $q^2$ for each $n_p$ is labeled by $q_{n_p}^2$ in the following.

\begin{table}[t!]
\caption{Momentum transfer squared $q^2$ in each $n_p = (Lp/2\pi)^2$.
$\nu_p$ is the number of the momentum assignment in the calculation
of $C^{K\pi}_{\mu,{\rm BC}}(\vec{p},t)$.}
\label{tab:q2}
\begin{tabular}{cccccccc}\hline\hline
& $q_0^2$ & $q^2_1$ & $q^2_2$ & $q^2_3$ & $q^2_4$ & $q^2_5$ & $q^2_6$ \\\hline
$n_p$ & 0 & 1 & 2 & 3 & 4 & 5 & 6\\
$\nu_p$ & 1 & 6 & 12 & 8 & 6 & 9 & 9 \\
$q^2$[GeV$^2$] & $-0.13103(48)$ & $-0.08980(33)$ & $-0.05656(25)$ & $-0.02792(20)$ & $-0.00239(17)$ & 0.02087(15) & 0.04239(13)\\
\hline\hline
\end{tabular}
\end{table}

We vary the time separation between the source and sink operators, 
$t_{\rm sep} = t_f - t_i = 36, 42$, and 48,
corresponding to 3.1, 3.6, and 4.1 fm in the physical unit, 
to study the excited state contributions of $R_\mu^{\rm BC}(\vec{p},t)$
in Eq.~(\ref{eq:def_Rmu}).
Since the statistical error increases for larger $t_{\rm sep}$,
the number of the random source $N_r = 2$ is 
chosen in the $t_{\rm sep} = 42$ and 48 cases,
while $N_r = 1$ in $t_{\rm sep} = 36$.

In order to increase statistics effectively,
on each configuration we perform the measurements with 8 different $t_i$
equally separated by 16 time separation,
4 temporal directions by rotating the configuration, 
and also average $C^{K\pi}_{\mu,{\rm BC}}(\vec{p},t)$ with 
its backward 3-point function calculated in $t_f \le t \le t_i$
with the same $t_{\rm sep}$.
In total the numbers of the measurements are
2560 for $t_{\rm sep} = 36$ 
and 5120 for $t_{\rm sep} = 42$ and 48, where 
the different choice of $N_r$ explained above is included. 
The statistical errors for all the observables are 
evaluated by the jackknife method with the bin size of 10 trajectories.

\section{$K_{l3}$ form factors}
\label{sec:form_factors}

In this section we discuss the two kinds of excited state contributions
in the ratio of the 3-point function $R_\mu^{\rm BC}(\vec{p},t)$
as explained in Sec.~\ref{sec:method}, 
which are the wrapping around effect
and the radial excited state contributions.
We also present the results for the $K_{l3}$ form factors, 
$f_+(q^2)$ and $f_0(q^2)$.
In the following discussions, 
we choose $t_i = 0$ so that $t_{\rm sep} = t_f$.

\subsection{Wrapping around effect}

\begin{figure}[ht!]
\makebox[.5\textwidth][r]{\includegraphics*[scale=.5]{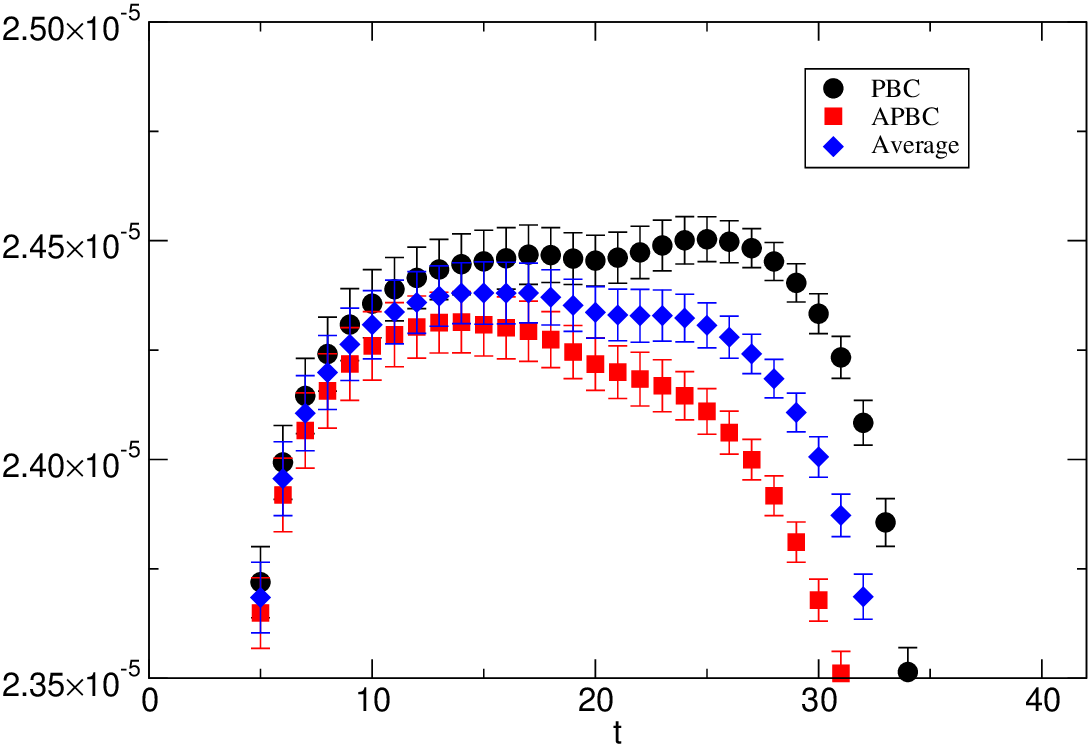}}
\caption{$t$ dependences for $R_{4}^{\rm PBC}(\vec{p},t)$ (circle), $R_{4}^{\rm APBC}(\vec{p},t)$ (square), and $R_{4}(\vec{p},t)$ (diamond) at $q^2 = q^2_0$ with $t_{\rm sep} = 42$.
}
\label{fig:wrap_eff_v4_q0}
\end{figure}

A typical example of the wrapping around effect in 
$R^{\rm PBC}_4(\vec{p},t)$ and $R^{\rm APBC}_4(\vec{p},t)$
defined in Eq.~(\ref{eq:def_Rmu})
is shown in Fig.~\ref{fig:wrap_eff_v4_q0},
where the ratios with $t_{\rm sep} = 42$ at $q^2 = q^2_0$ are plotted.
A clear discrepancy between $R^{\rm PBC}_4(\vec{p},t)$ and 
$R^{\rm APBC}_4(\vec{p},t)$ is observed in the region of $t > t_{\rm sep}/2$,
where the first term in the right hand side of 
Eq.~(\ref{eq:def_wrappind_around_effect}) is expected to have 
a large contribution.
The averaged ratio,
\begin{equation}
R_\mu(\vec{p},t) = \frac{R^{\rm PBC}_\mu(\vec{p},t) + R^{\rm APBC}_\mu(\vec{p},t)}{2},
\label{eq:average_R}
\end{equation}
has a milder $t$ dependence than the two ratios,
because the wrapping around effect $\Delta B_\mu(\vec{p},t)$ in Eq.~(\ref{eq:Rmu_time_dep}) cancels in the average.

\begin{figure}[ht!]
\makebox[.5\textwidth][r]{\includegraphics*[scale=.5]{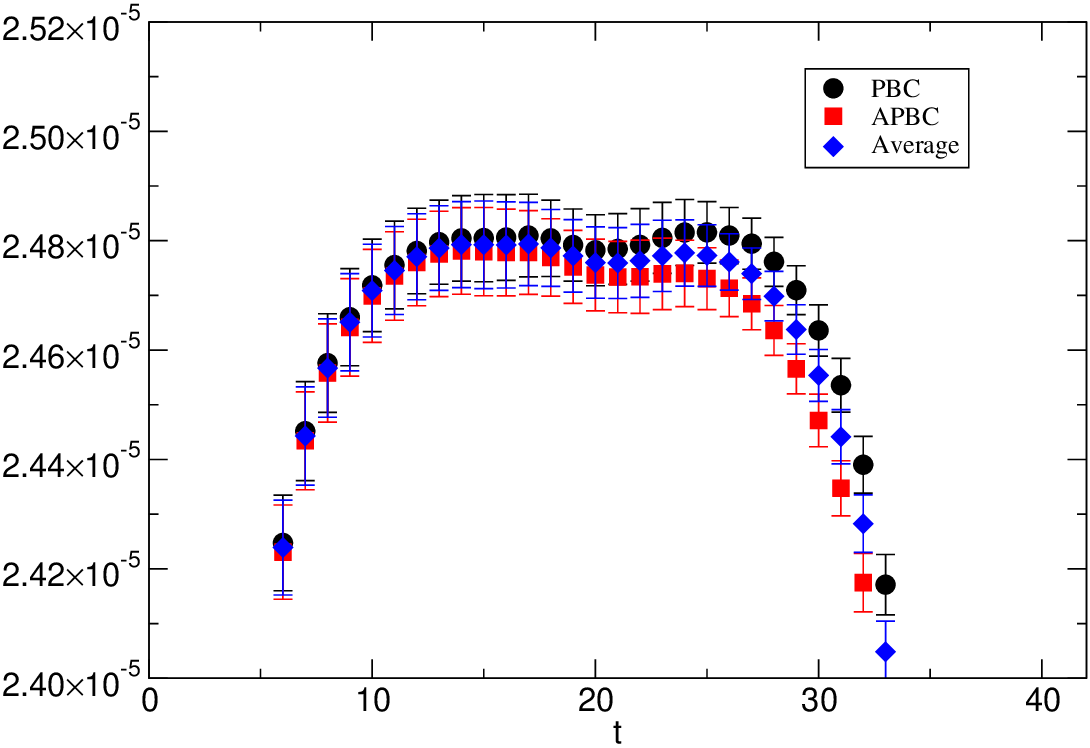}}
\caption{Same as Fig.~\ref{fig:wrap_eff_v4_q0}, but at $q^2 = q^2_1$.}
\label{fig:wrap_eff_v4_q1}
\end{figure}

Since the effect decreases as $p^2$ increases expected 
from Eq.~(\ref{eq:def_wrappind_around_effect}),
the discrepancy between the two ratios becomes smaller 
at $q^2 = q^2_1$ as shown in Fig.~\ref{fig:wrap_eff_v4_q1}.
Although the effect is small in large $p^2$,
we always adopt the averaged ratio $R_\mu(\vec{p},t)$ in the following analyses.

\subsection{$t_{\rm sep}$ dependence}
\label{sec:tsep}

\begin{figure}[ht!]
\makebox[.5\textwidth][r]{\includegraphics*[scale=.5]{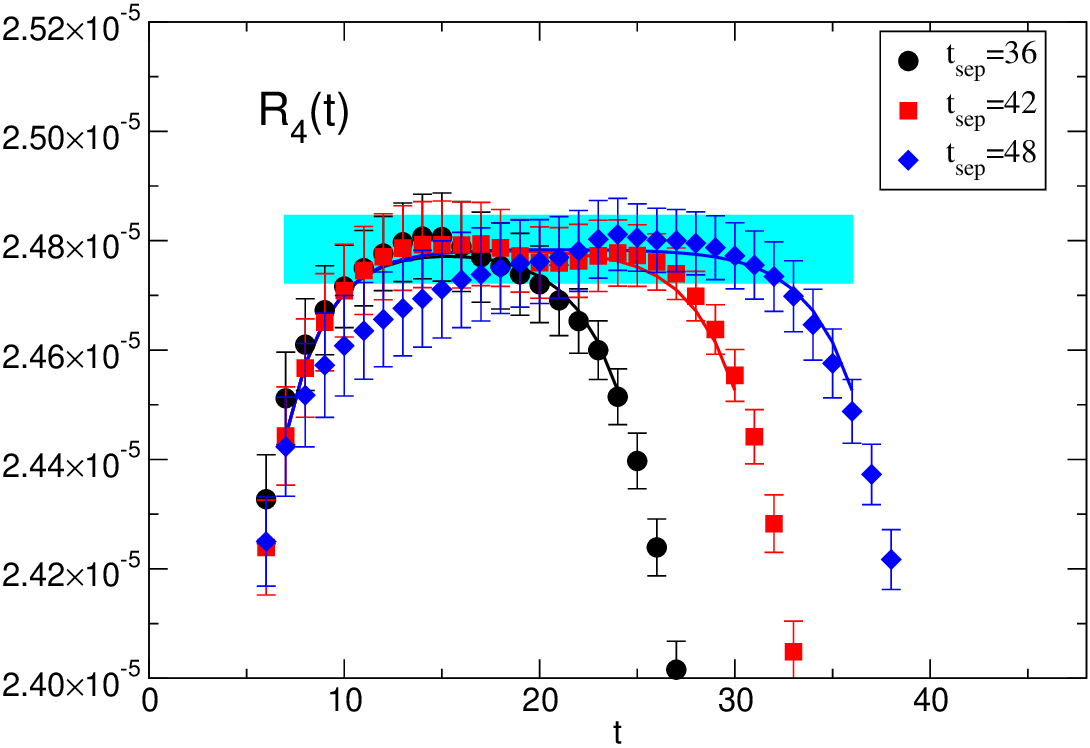}}
\makebox[.5\textwidth][r]{\includegraphics*[scale=.5]{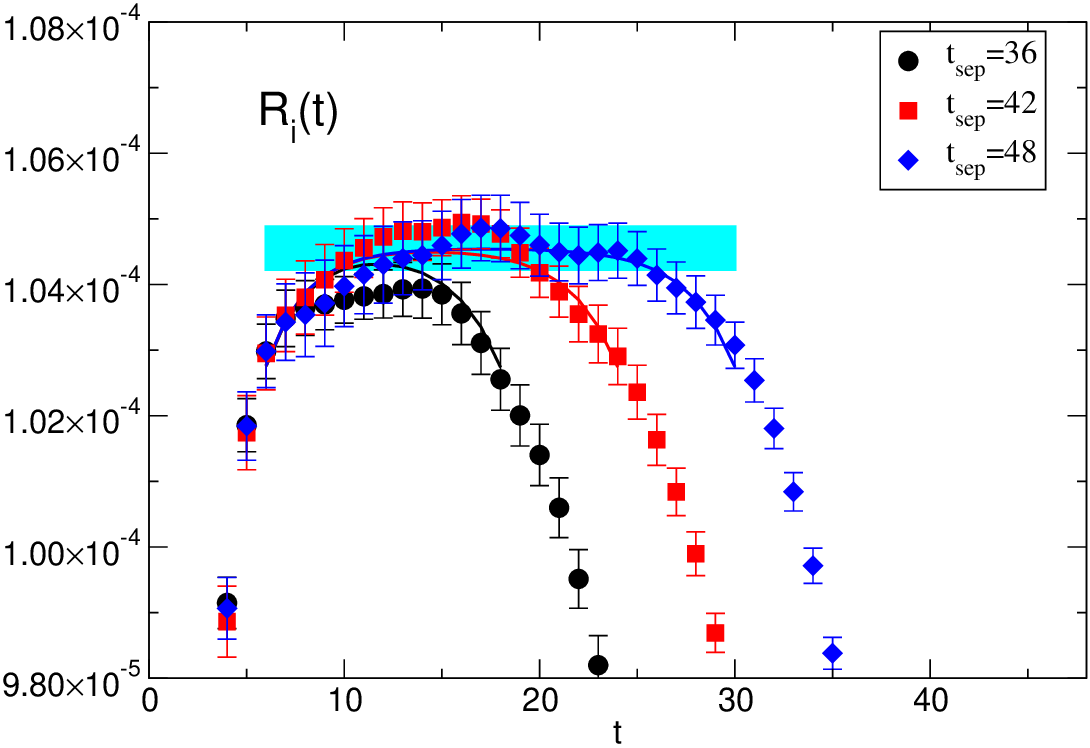}}
\caption{$t$ dependences for $R_4(t)$ (top) and $R_i(t)$ (bottom) at $q^2 = q^2_1$ with $t_{\rm sep} = 36$ (circle), 42 (square) and 48 (diamond), respectively. 
Fit curves with the fit form of Eq.~(\ref{eq:fit_form_Rmu}) with
the experimental $m_\pi^\prime$ and $m_K^\prime$ are also plotted.
The shaded band corresponds to the fit result of $R_\mu(p)$
with the one standard error,
and the $t$ region of the band expresses the fit range of $t_{\rm sep}=48$ data.}
\label{fig:tsep_extrapolation_q1}
\end{figure}

Figure~\ref{fig:tsep_extrapolation_q1} shows $t_{\rm sep}$ dependence of 
the averaged ratio $R_\mu(\vec{p},t)$ in Eq.~(\ref{eq:average_R}) at $q^2 = q_1^2$.
For $R_4(\vec{p},t)$, we observe a reasonable consistency
of the data with the different $t_{\rm sep}$
in flat regions between the source at $t=0$ and
sink at $t=t_{\rm sep}$.
For $R_i(\vec{p},t)$, flat regions
are shorter than those of $R_4(\vec{p},t)$,
and their shapes are non-symmetric.
In contrast to $R_4(\vec{p},t)$ the central values of 
$R_i(\vec{p},t)$ in the flat region of $t = $10--15
with $t_{\rm sep}=36$ are about 1\% smaller than 
those with $t_{\rm sep}=42$ and 48.
We consider that it is caused by excited state contributions
in $R_i(\vec{p},t)$,
and at first assume that it is the radial excitation of the mesons
as explained in Sec.~\ref{sec:method}.

To remove the contribution
and extract the matrix element 
$\langle \pi (\vec{p}) \left | V_{\mu} \right | K(\vec{0}) \rangle$
corresponding to the constant part in $R_\mu(\vec{p},t)$,
we fit $R_\mu(\vec{p},t)$ with a fit form given by,
\begin{eqnarray}
R_\mu(\vec{p},t) &=& R_\mu(p) 
+ \widetilde{A}^\pi_\mu(p)e^{-(E^\prime_\pi(p)-E_\pi(p))t}
+ \widetilde{A}_\mu^K(p) e^{-(m_K^\prime-m_K)(t_{\rm sep}-t)} ,
\label{eq:fit_form_Rmu}
\end{eqnarray}
where $R_\mu(p)$, $ \widetilde{A}^\pi_\mu(p)$, and $\widetilde{A}_\mu^K(p)$
are fit parameters,
and $E^\prime_\pi(p) = \sqrt{(m_\pi^\prime)^2 + p^2}$.

\begin{figure}[ht!]
\makebox[.5\textwidth][r]{\includegraphics*[scale=.5]{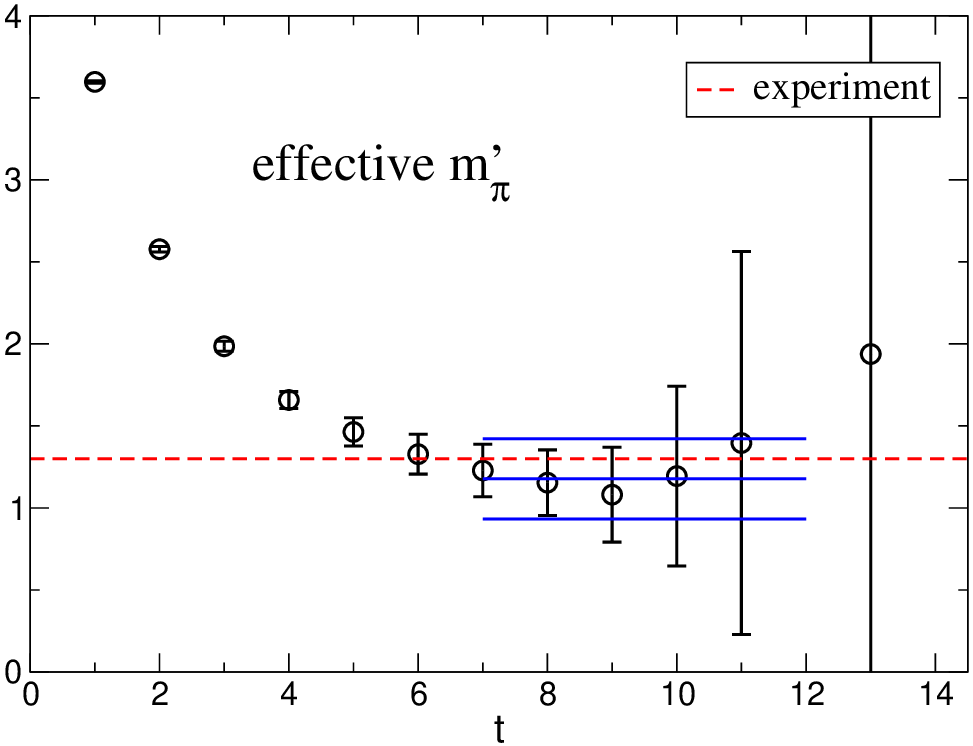}}
\makebox[.5\textwidth][r]{\includegraphics*[scale=.5]{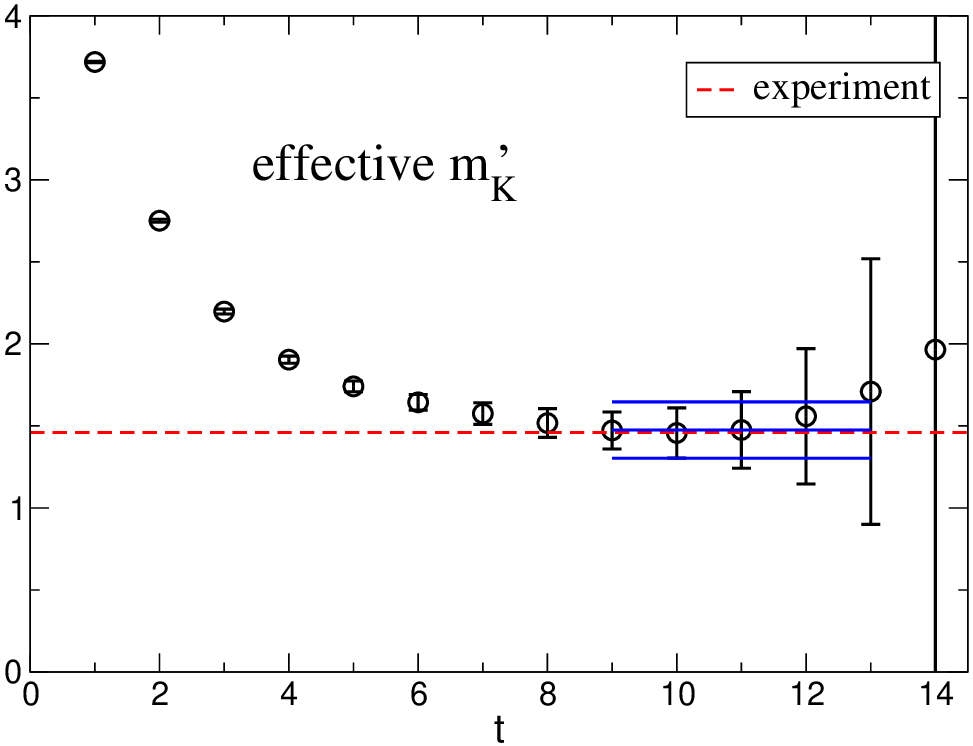}}
\caption{Effective masses defined by Eqs.~(\ref{eq:eff_first_exc}) and (\ref{eq:2-pt_sub}) for the first excited states for $\pi$ (top) and $K$ (bottom).
The solid lines express the fit results of the correlator in Eq.~(\ref{eq:2-pt_sub}) with the one standard error. The $t$ region of the lines denotes the fit range.}
\label{fig:eff_excited_mass}
\end{figure}

Since our simulation is carried out at the physical point,
the masses for the radial excited mesons, $m_\pi^\prime$ and $m_K^\prime$, 
are fixed to the experimental values
$m_\pi^\prime = 1.3$ GeV and $m_K^\prime = 1.46$ GeV in
PDG18~\cite{Tanabashi:2018oca}.
We examine if these masses are appropriate in our calculation
by effective masses for the first excited states in 
the 2-point functions.
The effective mass is evaluated from $C^X(\vec{0},t)$ without 
the ground state contribution defined as
\begin{equation}
m^\prime_{X,{\rm eff}} = \log\left(\frac{\overline{C}^X(\vec{0},t)}{\overline{C}^X(\vec{0},t+1)}\right),
\label{eq:eff_first_exc}
\end{equation}
where
\begin{equation}
\overline{C}^X(\vec{0},t) = C^X(\vec{0},t) - 
\frac{Z_X^2}{2m_X} (e^{-m_X t} + e^{-m_X(2T-t)}) ,
\label{eq:2-pt_sub}
\end{equation}
with $Z_X$ and $m_X$ obtained from a fit
using the asymptotic form in Eq.~(\ref{eq:asym_form_two-pt}).
As shown in Fig.~\ref{fig:eff_excited_mass},
we observe that the effective masses and also the fit results
for the first excited states
show reasonable consistencies with these experimental values.

The simultaneous fit results using all the $t_{\rm sep}$ data
with the fit form in Eq.~(\ref{eq:fit_form_Rmu})
are presented for $\mu = 4$ and $i$ at $q_1^2$
in Fig.~\ref{fig:tsep_extrapolation_q1}.
We employ uncorrelated fits in this analysis,
because our statistics are not enough to determine the covariance matrix
precisely.
It should be noted that in the fits 
the correlations among the data at different time slices and 
in different $t_{\rm sep}$ are taken into account by the jackknife method.
Thus, the effect of the uncorrelated fit is only a smaller value of $\chi^2/$dof
than that of the correlated fit with the correct covariance matrix.
The minimum time slice $t_{\rm min}$ of the fit range is fixed for all
$t_{\rm sep}$, while the maximum time slice $t_{\rm max}$ is changed for
each $t_{\rm sep}$ as $t_{\rm max} = t_{\rm sep} - t_{\rm fit}$.
In the $q^2 = q^2_1$ case as shown in the figure, 
$(t_{\rm min},t_{\rm fit}) = (7,12)$ and $(6,18)$
are chosen for $\mu=4$ and $i$, respectively.
The fit result of $R_\mu(p)$ represented by the shaded band in
the figure
agrees with the data in the flat region with the larger $t_{\rm sep}$
in both cases.

\begin{figure}[ht!]
\makebox[.5\textwidth][r]{\includegraphics*[scale=.5]{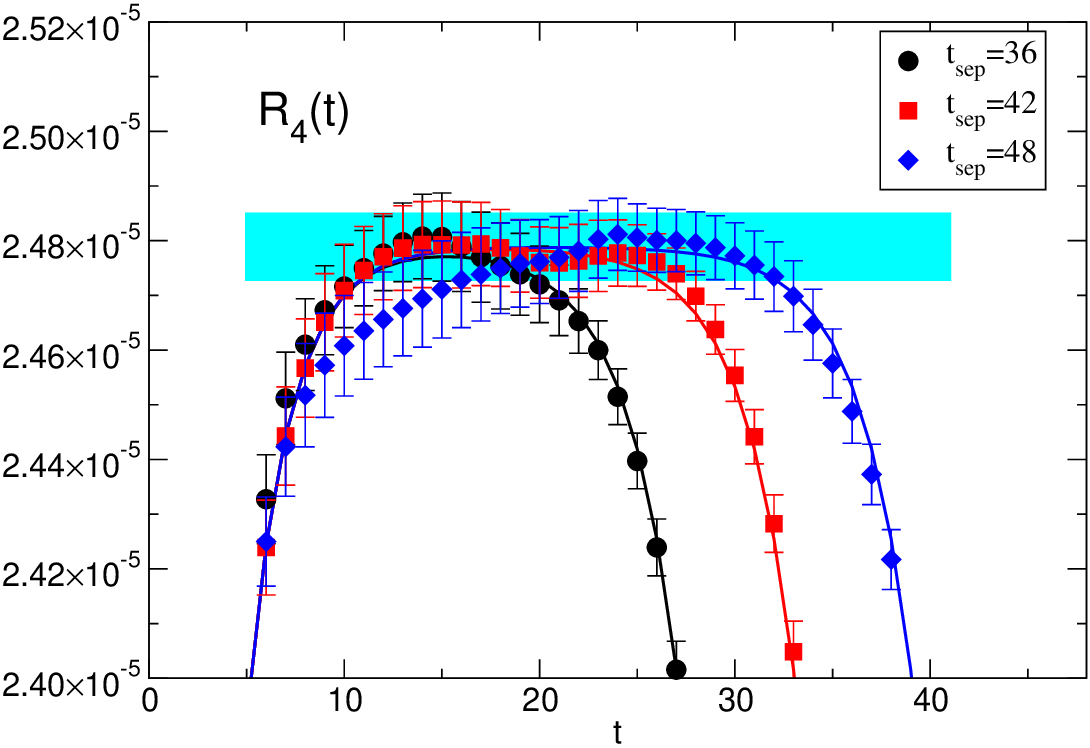}}
\makebox[.5\textwidth][r]{\includegraphics*[scale=.5]{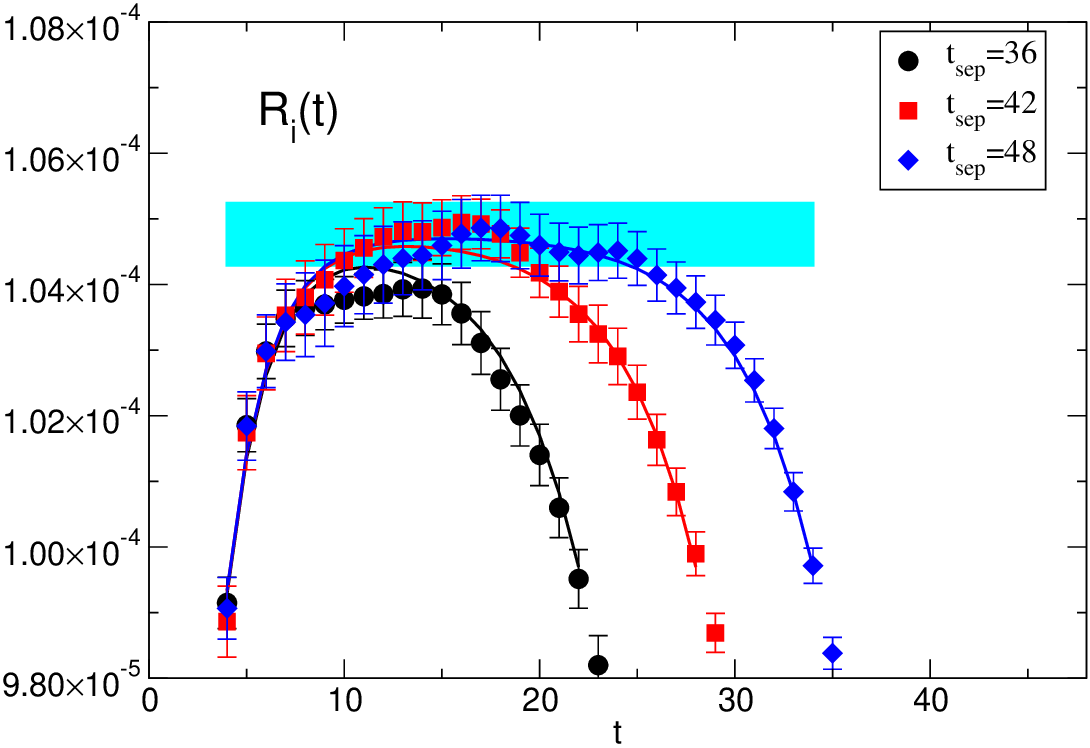}}
\caption{Same as Fig.~\ref{fig:tsep_extrapolation_q1},
but using fits with $E_\pi^\prime(p)$ and $m_K^\prime$
in Eq.~(\ref{eq:fit_form_Rmu}) as free parameters.}
\label{fig:tsep_extrapolation_q1_para}
\end{figure}

Although the above fit using the experimental $m_\pi^\prime$ and $m_K^\prime$
works well in our data, their contributions might not be  
the leading excited state contributions in the ratios.
In order to test the possibility,
we also fit $R_\mu(p)$ with $E_\pi^\prime(p)$ and $m_K^\prime$
as fit parameters in the fit form Eq.~(\ref{eq:fit_form_Rmu}),
and compare the results from the two analyses.
In this case we can choose wider fit ranges as 
$(t_{\rm min},t_{\rm fit}) = (5,7)$ and $(4,14)$ 
for $\mu = 4$ and $i$, respectively, than the ones with
the experimental $m_\pi^\prime$ and $m_K^\prime$.
The fit curves are presented in Fig.~\ref{fig:tsep_extrapolation_q1_para}.
The results of $R_\mu(p)$ agree with those in 
Fig.~\ref{fig:tsep_extrapolation_q1}, although the error of $R_i(p)$
becomes larger.

For later convenience, we call the data obtained from the fit with the fixed
$m_\pi^\prime$ and $m_K^\prime$ as ``A1'', and ones from 
another fit as ``A2''.
In the following, we use these two data to estimate a systematic error originating from the choice of the fitting form.
In each $q^2$, we carry out similar analyses to obtain
$R_\mu(p)$ for $\mu = 4$ and $i$, 
except for at $q^2 = q^2_0$ where only $R_4(p)$ is available.

We also perform the same analysis without the data of $t_{\rm sep}=36$
to study effects from the smallest $t_{\rm sep}$ data in our analysis.
It is found that the effect is not significant in our result,
because the fit result agrees with the above ones within the error.
Furthermore, we fit the data by adding 
a cross term of the second and third terms in Eq.~(\ref{eq:fit_form_Rmu}),
and also adding the second excited state contributions
corresponding to $m_{\pi^{(2)}} = 1.8$ GeV and 
$m_{K^{(2)}} = 1.86$ GeV. 
The results of $R_\mu(p)$
from those fits are statistically consistent with the ones in the above.

\subsection{Form factors}
\label{sec:form_factor}

\begin{figure}[ht!]
\makebox[.5\textwidth][r]{\includegraphics*[scale=.5]{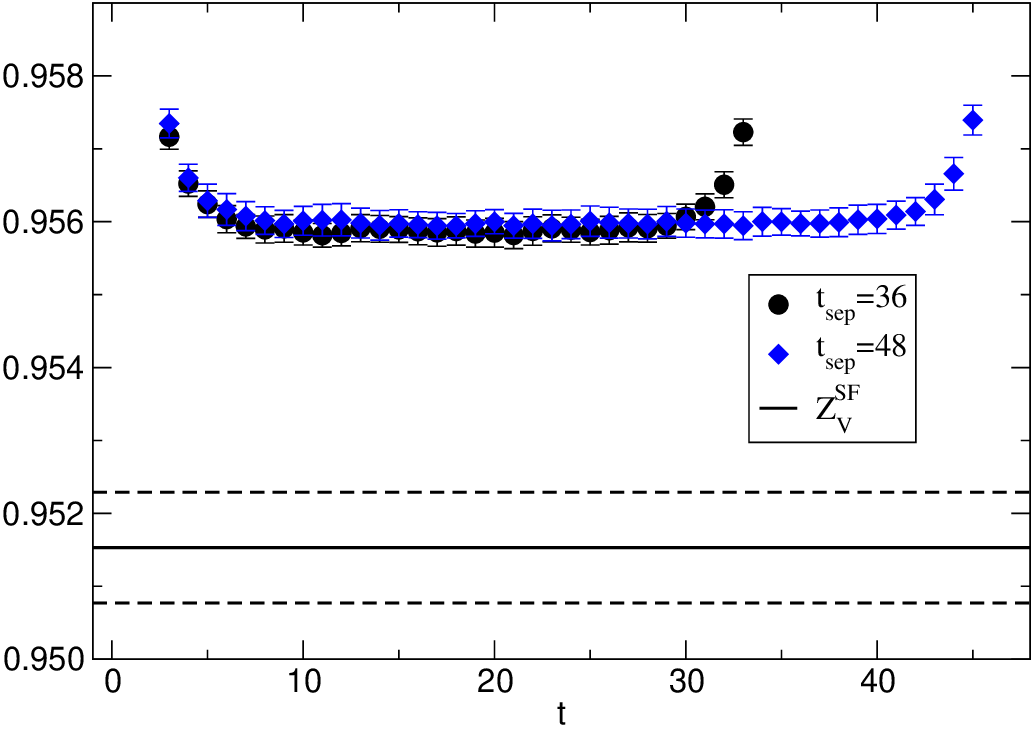}}
\caption{Renormalization factor of the vector current $R_{Z_V}(t)$ 
defined in Eq.~(\ref{eq:zv}) with $t_{\rm sep} = 36$ (circle) and 48 (square).
The solid and dashed lines represent the central value and error band of 
$Z_V^{\rm SF} = 0.95153(76)$ obtained by the SF scheme~\cite{Ishikawa:2015fzw}.
}
\label{fig:zv}
\end{figure}

For the renormalization of the local vector current,
the renormalization factor $Z_V$ is calculated from
the 3-point functions for $\pi$ and $K$ with the electromagnetic current
as presented in Eqs.~(\ref{eq:3-pt_em_pi}) and (\ref{eq:3-pt_em_K}).
To determine $Z_V$, a ratio $R_{Z_V}(t)$ is defined as
\begin{equation}
R_{Z_V}(t) = \sqrt{\frac{C^\pi(\vec{0},t_{\rm sep})C^K(\vec{0},t_{\rm sep})}
{C^{\pi\pi}_4(t)C^{KK}_4(t)}},
\label{eq:zv}
\end{equation}
whose value in a plateau region corresponds to $Z_V$.
Figure~\ref{fig:zv} shows that the data of $R_{Z_V}(t)$ 
with $t_{\rm sep} = 36$ and 48 agree with each other.
Thus, we determine $Z_V$ from a constant fit 
with $R_{Z_V}(t)$ of $t_{\rm sep} = 36$
in the middle $t$ region of $10 \le t \le 24$. 
The result of $Z_V = 0.95587(18)$ is 0.45\% larger than the value
obtained by the SF scheme~\cite{Ishikawa:2015fzw},
$Z_V^{\rm SF} = 0.95153(76)$, which is also shown in the figure.
From the discrepancy we will estimate a systematic error of the form factors 
in a later section.

\begin{figure}[ht!]
\makebox[.5\textwidth][r]{\includegraphics*[scale=.5]{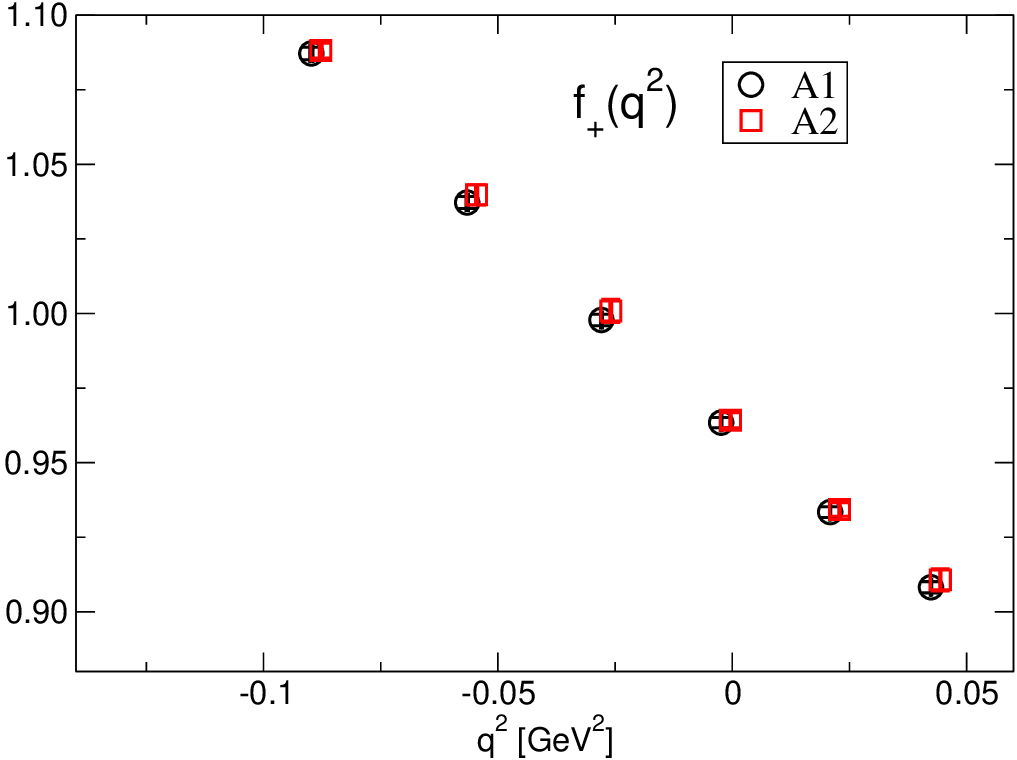}}
\makebox[.5\textwidth][r]{\includegraphics*[scale=.5]{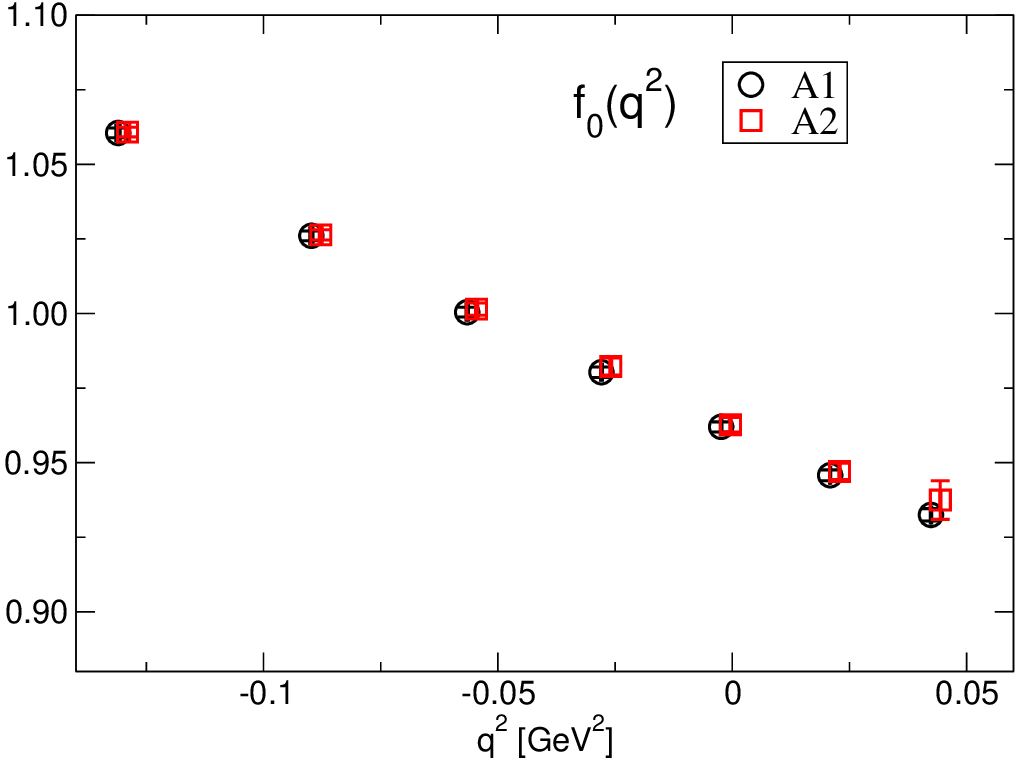}}
\makebox[.5\textwidth][r]{\includegraphics*[scale=.5]{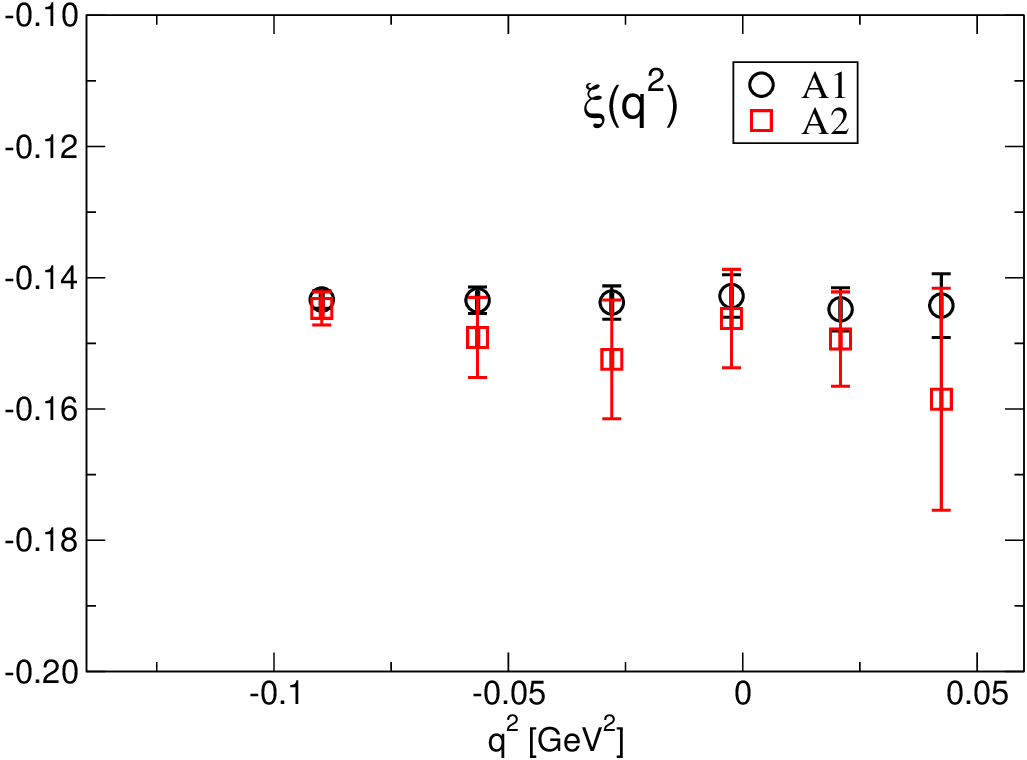}}
\caption{$K_{l3}$ form factors of $f_+(q^2)$ (top), $f_0(q^2)$ (middle), and
the ratio $\xi(q^2)$ (bottom) as a function of $q^2$.
Circle and square symbols represent A1 and A2 data sets, respectively.
The square symbols in the top and middle panels 
are slightly shifted in the $x$ direction for clarity.}
\label{fig:kl3_form_factor}
\end{figure}

Combining $Z_V$, $Z_\pi$ and $Z_K$ from the 2-point functions, and 
the results for $R_4(p)$ and $R_i(p)$,
we calculate the matrix elements 
$\langle \pi (\vec{p}) \left | V_4 \right | K(\vec{0}) \rangle$ and
$\langle \pi (\vec{p}) \left | V_{i} \right | K(\vec{0}) \rangle / p_i$, 
and then evaluate $f_+(q^2)$ and $f_0(q^2)$ with 
Eqs.~(\ref{eq:def_matrix_element}) and (\ref{eq:f0}) at each $q^2$, 
except at $q^2 = q^2_0$ where only $f_0(q^2_0)$ is obtained.
The form factors $f_+(q^2)$ and $f_0(q^2)$
obtained from the two data sets, A1 and A2,
are plotted in Fig.~\ref{fig:kl3_form_factor}
as a function of $q^2$
together with the ratio of the form factors $\xi(q^2)$
defined in Eq.~(\ref{eq:f0}).
Their numerical values are presented in Table~\ref{tab:kl3_form_factor}.
The results for $f_+(q^2)$ and $f_0(q^2)$ with the A1 data, which are taken to be the central values in our analysis, are obtained within less than 0.3\% statistical error.

\begin{table}[ht!]
\caption{Results for the form factors $f_+(q^2)$ and $f_0(q^2)$ together
with the ratio $\xi(q^2) = f_-(q^2)/f_+(q^2)$ defined in Eq.~(\ref{eq:f0})
at each $q^2$.
A1 and A2 data sets are explained in Sec.~\ref{sec:tsep}.}
\label{tab:kl3_form_factor}
\begin{tabular}{cccccccc}\hline\hline
& \multicolumn{3}{c}{A1} && \multicolumn{3}{c}{A2}\\
$q^2$ & $f_+(q^2)$ & $f_0(q^2)$& $\xi(q^2)$
&& $f_+(q^2)$ & $f_0(q^2)$& $\xi(q^2)$ \\\hline
$q^2_0$ & $\cdots$   & 1.0605(16) & $\cdots$      && $\cdots$   & 1.0608(16) & $\cdots$ \\
$q^2_1$ & 1.0872(21) & 1.0260(16) & $-0.1433(13)$ && 1.0881(25) & 1.0264(17) & $-0.1447(25)$ \\
$q^2_2$ & 1.0372(20) & 1.0004(16) & $-0.1434(20)$ && 1.0398(34) & 1.0015(20) & $-0.1491(61)$ \\
$q^2_3$ & 0.9978(19) & 0.9803(18) & $-0.1438(25)$ && 1.0009(39) & 0.9822(29) & $-0.1524(91)$ \\
$q^2_4$ & 0.9634(17) & 0.9620(17) & $-0.1428(32)$ && 0.9642(25) & 0.9627(25) & $-0.1462(75)$ \\
$q^2_5$ & 0.9334(18) & 0.9458(19) & $-0.1448(33)$ && 0.9343(23) & 0.9470(27) & $-0.1493(72)$ \\
$q^2_6$ & 0.9082(19) & 0.9325(21) & $-0.1442(49)$ && 0.9107(37) & 0.9375(65) & $-0.159(17)$ 
\\\hline\hline
\end{tabular}
\end{table}

\section{$q^2$ dependence of form factors}
\label{sec:interpolation}

It is necessary for the determination of $|V_{us}|$
to extract $f_+(0)$ from $f_+(q^2)$ and $f_0(q^2)$.
Although in our calculation $q^2_4$ is very close to zero,
we need a small interpolation using some fit function.
In this section we explain the interpolation procedures and give the results 
for $f_+(0)$ with systematic errors.
All the interpolations are carried out with uncorrelated fits
due to a lack of enough statistics to determine a precise covariance matrix.
We note that, as in the fits in Sec.~\ref{sec:tsep}, 
the correlation among the data is treated by the jackknife method,
so that the value of $\chi^2$/dof in this fit
can be smaller than the one in the correlated fit.
Furthermore, we also discuss the shape of the form factors and
the phase space integral evaluated with our form factors.

\subsection{Interpolations to $q^2 = 0$}
\label{sec:interpolations}

For the form factors, $f_+(q^2)$ and $f_0(q^2)$,
the next-to-leading order (NLO) formulae are available
in the SU(3) ChPT~\cite{Gasser:1984ux,Gasser:1984gg}.
We employ the following fit functions for the interpolations to $q^2 = 0$,
which are based on the NLO ChPT formulae:
\begin{eqnarray}
f_+(q^2) &=& 1 - \frac{4}{F_0^2} L_9 q^2 + K_+(q^2) + c_0 + c_2^+ q^4 ,
\label{eq:NLO_chpt_f+}\\
f_0(q^2) &=& 1 - \frac{8}{F_0^2} L_5 q^2 + K_0(q^2) + c_0 + c_2^0 q^4 ,
\label{eq:NLO_chpt_f0}
\end{eqnarray}
where $F_0$ is the pion decay constant\footnote{We adopt the normalization of $F_\pi \sim 132$ MeV at the physical point.} in the chiral limit, and
$L_9$, $L_5$, $c_0$, and $c^{+,0}_{2}$ are free parameters.
The constraint of $f_+(0) = f_0(0)$ is required,
so that the same $c_0$ appears in the two fit functions.
The functions $K_+(q^2)$ and $K_0(q^2)$ are given in 
Appendix~\ref{app:chpt_formulae}, which depend on $m_\pi$, $m_K$, $q^2$,
$F_0$, and the scale $\mu$.
In our analyses we fix $\mu = 0.77$ GeV.
The last two terms in each fit function
can be regarded as a part of the NNLO analytic terms in the SU(3) ChPT.
In this point of view, the constant term $c_0$ represents a sum of
$m_\pi^4$, $m_K^4$, and $m_\pi^2 m_K^2$ terms,
because $m_\pi^2$ and $m_K^2$ are constant in our analysis
due to the physical point calculation.

\begin{figure}[ht!]
\makebox[.5\textwidth][r]{\includegraphics*[scale=.5]{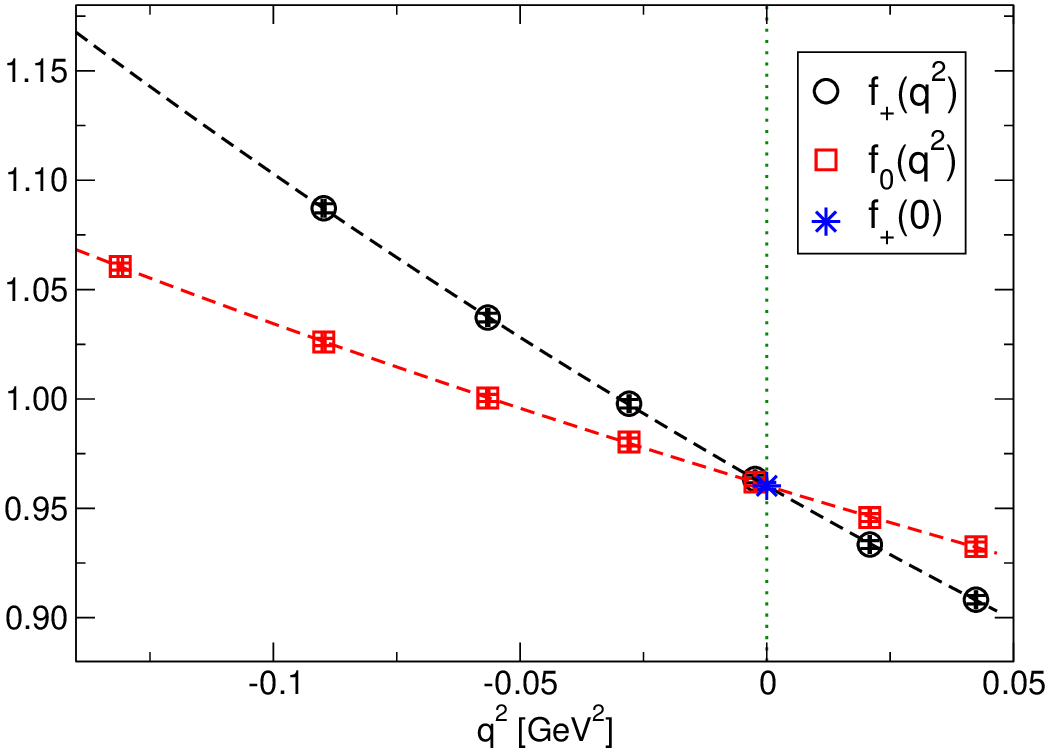}}
\makebox[.5\textwidth][r]{\includegraphics*[scale=.5]{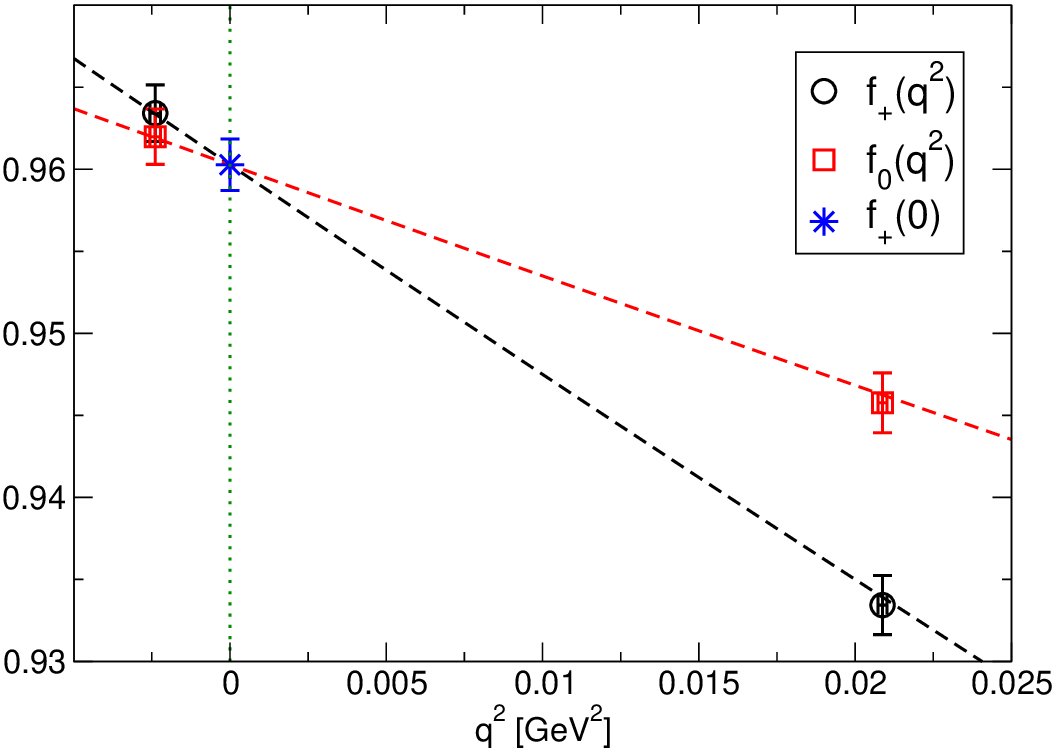}}
\caption{Interpolation of $K_{l3}$ form factors
with the fit forms based on the NLO SU(3) ChPT formulae in 
Eqs.~(\ref{eq:NLO_chpt_f+}) and (\ref{eq:NLO_chpt_f0}) with the fixed $F_0$
using the A1 data.
The top and bottom panels present the fit results in all $q^2$ regions
we calculated and the ones near the $q^2=0$ region, respectively.
The cross expresses the fit result of $f_+(0)$.
}
\label{fig:q2_interpolation_su3_chpt}
\end{figure}

In an interpolation we fix $F_0 = 0.11205$ GeV, which is determined from  
the average of the ratios,
$F/F_0 = 1.229(59)$~\cite{Allton:2008pn} and 
$F/F_0 = 1.078(44)$~\cite{Aoki:2008sm},
summarized in the FLAG review 2019~\cite{Aoki:2019cca},
and using $F = 0.12925$ GeV from our SU(2) ChPT analysis~\cite{Kakazu:2017fhv}.
$F$ is the pion decay constant of the SU(2) ChPT
in the chiral limit.
We carry out a simultaneous fit using
all the A1 data for $f_+(q^2)$ and $f_0(q^2)$ including $f_0(q^2_0)$.
The fit results are presented in Fig.~\ref{fig:q2_interpolation_su3_chpt}.
The top panel 
shows that the fit works well in all the $q^2$ region for both the
form factors.
The interpolated result of $f_+(0) = 0.9603(16)$ has a comparable error with
the nearest data to $q^2 = 0$ as shown in the bottom panel, 
which is an enlarged figure
of the top panel near the $q^2 = 0$ region.
The values for the fit results are tabulated in Table~\ref{tab:fit_chpt}.
It is noted that the validity of the constraint 
$f_+(0) = f_0(0)$ in the fit is confirmed 
by the fact that the independent fit results for $f_+(0)$ and $f_0(0)$ 
agree with each other. They are also consistent with the simultaneous fit 
result in the above.

We also carry out another fit using the same fit forms of
Eqs.~(\ref{eq:NLO_chpt_f+}) and (\ref{eq:NLO_chpt_f0}),
while setting $F_0$ as a free parameter and $c_0 = 0$.
This fit result of $f_+(0)$ is consistent with 
the one obtained from the above fit.
The fit result of $F_0 = 0.1006(20)$ GeV is compatible to the one
assumed in the above fit.
This observation indicates that a systematic error 
due to the fixed $F_0$ should be small in the above fit.
The fit results are summarized in Table~\ref{tab:fit_chpt}. 
The table also contains the fit results using the A2 data.
The results of $f_+(0)$ with the A2 data agree with
the ones with the A1 data,
while they have larger errors than those with the A1 data.
Our results for $L_9$ and $L_5$ show similar values to
the previous lattice QCD results for the low energy constants in the SU(3) ChPT 
summarized in the FLAG review 2019~\cite{Aoki:2019cca},
{\it i.e.}, $L_9 \sim (2.4$--3.8)$\times 10^{-3}$ and
$L_5 \sim (0.9$--1.5)$\times 10^{-3}$.
Note that these values are given in the chiral limit 
for all the quark masses,
so that they cannot be directly compared with our results obtained 
at the physical quark masses.

We also employ several different fit forms for the interpolation, 
such as a mono-pole function,
a simple quadratic function of $q^2$, 
and variations of the $z$-parameter expansion~\cite{Bourrely:2008za}.
The fit forms and the fit results are 
summarized in Appendix~\ref{app:q2_fit_result}.
The results of $f_+(0)$ obtained from these fits 
are also consistent with the ones obtained from the above ChPT analyses.
Furthermore, we confirm that the result is not changed,
when the fit range of $q^2$ is squeezed as $q^2_2 \le q^2 \le q_5^2$
in the NLO ChPT fit with the fixed $F_0$ using the A1 data, 
which gives $f_+(0) = 0.9604(16)$.
Based on these fit analyses we conclude that 
the systematic error originating from the fit form dependence for the interpolation is as small as the statistical error in
our result of $f_+(0)$.

\begin{table}[ht!]
\caption{Fit results of $K_{l3}$ form factors
based on the NLO SU(3) ChPT formulae in 
Eqs.~(\ref{eq:NLO_chpt_f+}) and (\ref{eq:NLO_chpt_f0})
together with the value of the uncorrelated $\chi^2/$dof.
A1 and A2 data sets are explained in Sec.~\ref{sec:tsep}.
fit-1 and fit-2 denote fits with the fixed and free $F_0$, respectively,
as explained in Sec.~\ref{sec:interpolations}.
We also list the results for $f_+(0)$, slope, curvature, and 
phase space integral.
}
\label{tab:fit_chpt}
\begin{tabular}{cccccc}\hline\hline
& \multicolumn{2}{c}{A1} && \multicolumn{2}{c}{A2}\\
& fit-1 & fit-2 && fit-1 & fit-2 \\\hline
$L_9$ [$10^{-3}$] & 3.924(57) & 3.14(14) && 3.94(11) & 3.27(25) \\
$L_5$ [$10^{-4}$] & 6.94(28) & 4.88(41) && 6.73(52) & 5.01(55) \\
$c_2^+$ [GeV$^{-4}$] & 1.19(17) & 1.15(17) && 1.13(36) & 1.10(36) \\
$c_2^0$ [GeV$^{-4}$] & $-0.40(11)$ & $-0.65(12)$ && $-0.36(19)$ & $-0.57(21)$ \\
$c_0$ & $-0.0077(16)$ & $\cdots$ && $-0.0063(24)$ & $\cdots$ \\
$F_0$ [GeV] & $\cdots$ & 0.1007(20)&& $\cdots$ & 0.1024(32) \\
$\chi^2/$dof & 0.05 & 0.05 && 0.18 & 0.18 \\
$f_+(0)$ & 0.9603(16) & 0.9603(16) && 0.9616(24) & 0.9617(24) \\
$\lambda_+^\prime$ [$10^{-2}$] & 2.618(37) & 2.635(37) && 2.627(70) & 2.643(70) \\
$\lambda_0^\prime$ [$10^{-2}$] & 1.384(37) & 1.393(37) && 1.355(68) & 1.365(69) \\
$\lambda_+^{\prime\prime}$ [$10^{-3}$] & 1.06(13) & 1.07(13) && 1.01(29) & 1.02(29) \\
$\lambda_0^{\prime\prime}$ [$10^{-3}$] & 0.401(91) & 0.381(95) && 0.43(15) & 0.41(17) \\
$I^e_{K^0}$ & 0.15481(13) & 0.15481(13) && 0.15482(22) & 0.15482(22) \\
$I^\mu_{K^0}$ & 0.10249(12) & 0.10248(12) && 0.10244(21) & 0.10243(21)\\
$f_+(0)\sqrt{I^e_{K^0}}$ & 0.37783(62) & 0.37784(62) && 0.37837(93) & 0.37837(93) \\
$f_+(0)\sqrt{I^\mu_{K^0}}$ & 0.30742(49) & 0.30742(49) && 0.30777(68) & 0.30777(68)
\\\hline\hline
\end{tabular}
\end{table}

\subsection{Result of $f_+(0)$}
\label{sec:result_f0}

From the fit results discussed in the previous subsection,
whose values are tabulated in Table~\ref{tab:fit_chpt} and
Tables in Appendix~\ref{app:q2_fit_result},
we obtain the result of $f_+(0)$ as
\begin{equation}
f_+(0) = 0.9603(16)(^{+14}_{\ -4})(44)(19)(1),
\label{eq:central_res_f00}
\end{equation}
where the central value and statistical error (the first error)
are determined from the fit result based on the ChPT 
formulae in Eqs.~(\ref{eq:NLO_chpt_f+}) and (\ref{eq:NLO_chpt_f0})
with the fixed $F_0$ using the A1 data.
The second error is the systematic one for the fit form dependence,
which is estimated from the deviation of the various fit results,
tabulated in Table~\ref{tab:fit_chpt} and 
tables in Appendix~\ref{app:q2_fit_result},
from the central value.

The third error is the systematic one for the discrepancy of $Z_V$ and $Z_V^{\rm SF}$,
0.45\%, discussed in Sec.~\ref{sec:form_factor}.
We consider that it is regarded as the order of a systematic error 
due to the finite lattice spacing effect, 
because the discrepancy should vanish in the continuum limit.
In our calculation this value is larger than 0.19\%, which is 
an order estimation of a discretization error from the higher 
order contributions in the ChPT formula as
$(1-f_+(0))\times (\Lambda a)^2$ with $\Lambda = 0.5$ GeV.
Since $f_+(0) = 1$ is fixed from the symmetry in the LO ChPT,
$1-f_+(0)$ represents the higher order contributions.
This estimation was used in the
previous studies~\cite{Aoki:2017spo,Boyle:2013gsa,Boyle:2015hfa}.\footnote{$\Lambda = 0.3$ GeV was employed in Ref.~\cite{Boyle:2013gsa}.
We quote the ChPT estimation as the fourth error, because it comes
from a different effect of the discretization error rather than that in $Z_V$.}

We also estimate a systematic error of the isospin symmetry breaking effect
by replacing the NLO functions $K_+(q^2)$ and $K_0(q^2)$ in 
Eqs.~(\ref{eq:NLO_chpt_f+}) and (\ref{eq:NLO_chpt_f0})
by the ones for $f^{K^0\pi^-}_+$ and $f^{K^0\pi^-}_0$
in the NLO ChPT with the isospin breaking~\cite{Gasser:1984ux,Bijnens:2007xa}.
We evaluate $f^{K^0\pi^-}_+(0) = 0.9604$
using the fit parameters obtained from the fit with the fixed $F_0$ for the A1 data and
the experimental $\pi$ and $K$ masses\footnote{We use the $\pi^0$-$\eta$ mixing angle $\varepsilon = 0.0116$, which is estimated using the quark masses in PDG~\cite{Tanabashi:2018oca}.} in PDG~\cite{Tanabashi:2018oca}.
Comparing $f^{K^0\pi^-}_+(0)$ with $f_+(0)$, it is found that the effect is 
much smaller than other errors.
We quote their deviation as the fifth error in Eq.~(\ref{eq:central_res_f00}).
It is an important future work for a nonperturbative estimation of this error
to perform calculation including QED effect,
such as one in the $K_{l2}$ decay~\cite{DiCarlo:2019thl}.

We do not include the systematic error of the finite volume effect,
because our physical volume is large enough 
to suppress the effect.
The estimate based on ChPT, $(1-f_+(0))\times e^{-m_\pi L}$, gives 
0.002\%, which is much smaller than other errors.
In the following we will not discuss this systematic error.

\begin{figure}[ht!]
{\includegraphics*[scale=.5]{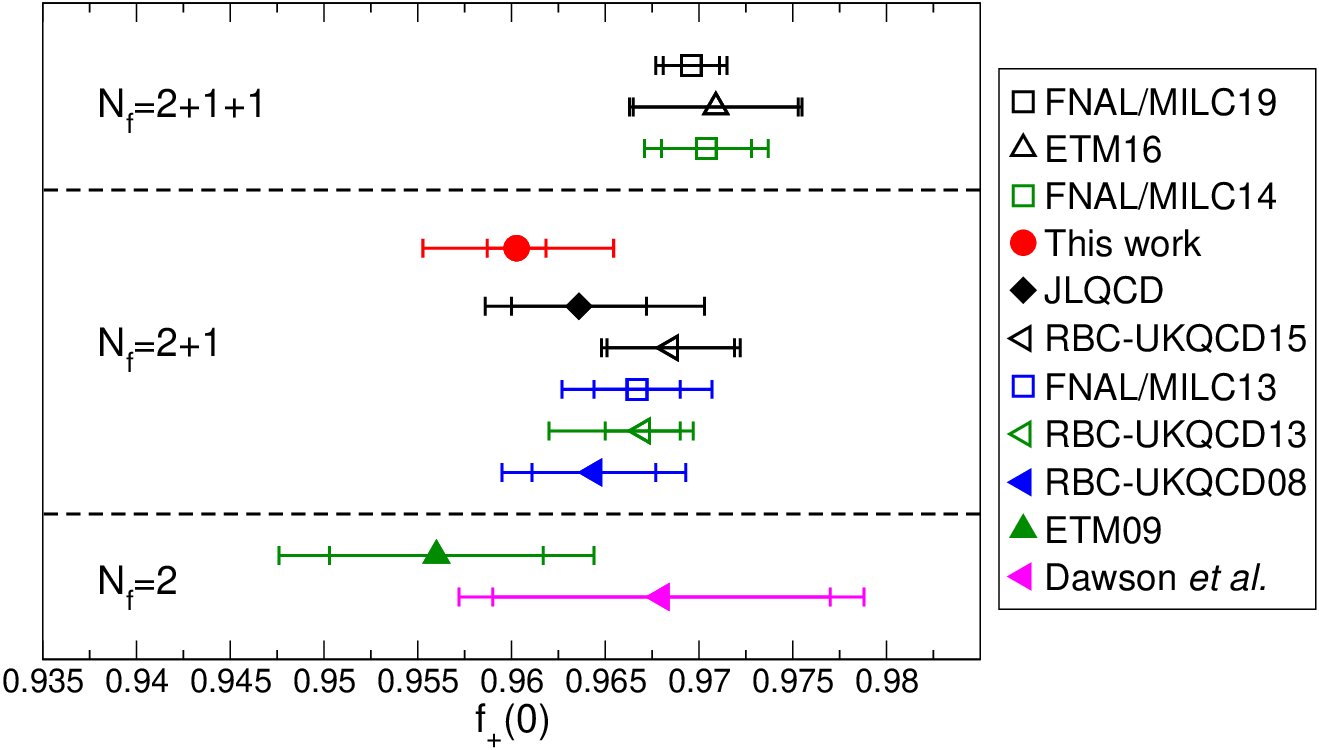}}
\caption{Comparison of $f_+(0)$ with the previous results
in dynamical quark calculations.
Our result is represented by the circle symbol.
Square, up triangle, diamond, and left triangle symbols denote
staggered~\cite{Bazavov:2012cd,Bazavov:2013maa,Bazavov:2018kjg}, twisted~\cite{Lubicz:2009ht,Carrasco:2016kpy}, overlap~\cite{Aoki:2017spo}, and
domain wall~\cite{Dawson:2006qc,Boyle:2007qe,Boyle:2013gsa,Boyle:2015hfa} quark calculations, respectively.
The filled symbols represent the results at a finite lattice spacing.
The inner and outer errors express the statistical and total errors.
The total error is evaluated by adding the statistical and systematic
errors in quadrature.}
\label{fig:comparison_f0_0}
\end{figure}

Figure~\ref{fig:comparison_f0_0} shows comparison of our result
with the previous dynamical lattice QCD calculations~\cite{Bazavov:2012cd,Boyle:2007qe,Boyle:2013gsa,Boyle:2015hfa,Aoki:2017spo,Bazavov:2013maa,Carrasco:2016kpy,Bazavov:2018kjg,Dawson:2006qc,Lubicz:2009ht}.
Our result is reasonably consistent with the previous $N_f = 2$~\cite{Dawson:2006qc,Lubicz:2009ht} and $N_f = 2+1$~\cite{Bazavov:2012cd,Boyle:2007qe,Boyle:2013gsa,Boyle:2015hfa,Aoki:2017spo} calculations,
while it is slightly smaller than the recent $N_f = 2+1+1$ results~\cite{Bazavov:2013maa,Carrasco:2016kpy,Bazavov:2018kjg}.
The largest discrepancy in comparison with the previous results is 1.7 $\sigma$ from the one
in Ref.~\cite{Bazavov:2018kjg} in the total error.
At present the reason of the discrepancy is not clear.
However, an analysis using only the physical point data in 
Ref.~\cite{Bazavov:2018kjg}
gives a smaller value than their result in the figure, so that
the discrepancy would become smaller 
with larger systematic errors~\cite{Bazavov:2018kjg}.
In order to understand the source of the discrepancy,
it is important to reduce our uncertainties, especially,
the finite lattice spacing effect, which is the largest error in
our calculation.
For this purpose, in the next step we will calculate the form factors
using other sets of PACS10 configurations 
with the finer lattice spacings at the physical point
to evaluate $f_+(0)$ in the continuum limit.

\subsection{Shape of form factors}

The slopes for the form factors are defined by 
the Taylor expansion in a vicinity of $q^2 = 0$ as,
\begin{equation}
f_s(q^2) = f_+(0) \left( 1 + \lambda_s^\prime 
\left(\frac{-q^2}{m_{\pi^-}^2}\right)
+ \lambda_s^{\prime\prime} \left(\frac{-q^2}{m_{\pi^-}^2}\right)^2 
+ \cdots \right) ,
\end{equation}
where $s = +$ and 0, and $m_{\pi^-} = 0.13957061$ GeV.

\begin{figure}[ht!]
\makebox[.5\textwidth][r]{\includegraphics*[scale=.5]{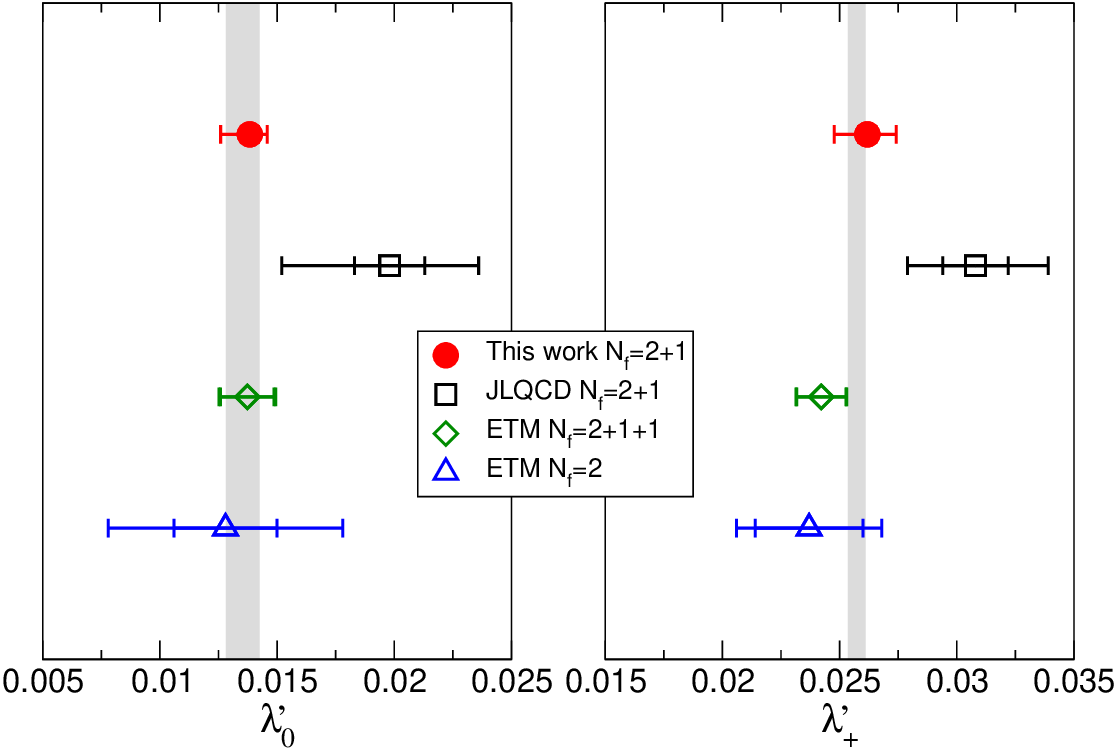}}
\caption{Comparison for the slopes of the form factors,
$\lambda_0^\prime$ and $\lambda_+^\prime$, with 
previous lattice QCD results~\cite{Aoki:2017spo,Lubicz:2009ht,Carrasco:2016kpy}.
The experimental results~\cite{Moulson:2017ive} are denoted
by the shaded bands.
The inner and outer errors express the statistical and total errors.
The total error is evaluated by adding the statistical and systematic
errors in quadrature.}
\label{fig:comparison_lamda}
\end{figure}

The fit of the form factors discussed in Sec.~\ref{sec:interpolations}
gives the slope $\lambda_s^\prime$ and curvature 
$\lambda_s^{\prime\prime}$,
whose results are presented in Table~\ref{tab:fit_chpt} and
Tables in Appendix~\ref{app:q2_fit_result}.
For $\lambda_s^\prime$, we obtain
\begin{eqnarray}
\lambda_+^\prime &=& 2.618(37)(^{+26}_{-68})(118)(5) \times 10^{-2}, \\
\lambda_0^\prime &=& 1.384(37)(^{+20}_{-93})(62)(5) \times 10^{-2},
\end{eqnarray}
where the central value, the first and second errors
are determined in a similar way to the $f_+(0)$ case shown in
Sec.~\ref{sec:result_f0}.
Since the systematic error coming from $Z_V$ affects only
the overall constant $f_+(0)$, there is no corresponding systematic error
for $\lambda_s^\prime$ and $\lambda_s^{\prime\prime}$.
The third and fourth errors are discretization effects estimated from 
the higher order in ChPT and the isospin symmetry breaking effect evaluated
as in $f_+(0)$, respectively.
Note that since the slopes and curvatures originate from
the higher order contributions in ChPT,
the corresponding discretization errors, 5\%, are much larger than the one 
in $f_+(0)$, 0.19\%.
Those results are well consistent with 
the experimental ones~\cite{Moulson:2017ive}\footnote{$\lambda_0^\prime$ is evaluated using $\lambda_0^\prime = (m_{\pi^-}^2/\Delta_{K\pi})(\log C - 0.0398(44))$~\cite{Bernard:2009zm} with $\Delta_{K\pi} = m_{K^0}^2 - m_{\pi^-}^2$ 
and $\log C = 0.1985(70)$ and $m_{K^0} = 0.497611$ GeV.},
$\lambda_+^\prime = 2.575(36) \times 10^{-2}$ and
$\lambda_0^\prime = 1.355(71) \times 10^{-2}$,
and the previous lattice ones~\cite{Aoki:2017spo,Lubicz:2009ht,Carrasco:2016kpy} as shown in Fig.~\ref{fig:comparison_lamda}.

For $\lambda_s^{\prime\prime}$, we obtain
\begin{eqnarray}
\lambda_+^{\prime\prime} &=& 1.06(13)(^{+32}_{-16})(5)(0) \times 10^{-3}, \\
\lambda_0^{\prime\prime} &=& 0.40(9)(^{+26}_{\ -8})(2)(0) \times 10^{-3},
\end{eqnarray}
where the results and errors are determined in a similar way to
$\lambda_s^{\prime}$.
For the curvatures, the isospin symmetry breaking effects are negligible in our precision.
Those results agree with
the experimental ones calculated with the dispersive representation~\cite{Bernard:2009zm}\footnote{
$\lambda_s^{\prime\prime}$ is expressed by $\lambda_s^{\prime}$
in the dispersive representation~\cite{Bernard:2009zm} as,
$\lambda_+^{\prime\prime} = (\lambda_+^\prime)^2 + (5.79^{+1.91}_{-0.97}) \times 10^{-4}$ and
$\lambda_0^{\prime\prime} = (\lambda_0^\prime)^2 + (4.16\pm 0.56) \times 10^{-4}$.},
$\lambda_+^{\prime\prime} = 1.24(^{+19}_{-10}) \times 10^{-3}$ and
$\lambda_0^{\prime\prime} = 0.600(59) \times 10^{-3}$,
and also an average of the experimental results~\cite{Antonelli:2010yf},
$\lambda_+^{\prime\prime} = 1.57(48) \times 10^{-3}$.


\subsection{Phase space integral}
\label{sec:phase_space_integral}

Since our results for the slopes and the curvatures of the form factors
agree with the experiment,
we evaluate the phase space integral, $I^l_K$ in Eq.~(\ref{eq:decay_width_kl3}),
which is usually calculated using
the $q^2$ dependence of the experimental form factors.
The phase space integral~\cite{Leutwyler:1984je} is given by
\begin{equation}
I^l_K = \int^{t_{\rm max}}_{m_l^2} dt
\frac{\lambda^{3/2}}{M_K^8}
\left( 1 + \frac{m_l^2}{2 t} \right)
\left( 1 - \frac{m_l^2}{t} \right)^2
\left(
\overline{F}_+^2(t) + 
\frac{3 m_l^2 \Delta_{K\pi}^2}{(2t + m_l^2)\lambda}\overline{F}_0^2(t)
\right),
\end{equation}
where $\lambda = ( t - \Sigma ) ( t - t_{\rm max} )$ with
$\Sigma = ( M_K + M_\pi )^2$ and $t_{\rm max} = ( M_K - M_\pi )^2$,
$\overline{F}_s(t) = f_s(-t)/f_+(0)$ with $s = +$ and 0,
$t = -q^2$, and $m_l$ is the mass of the lepton $l$.
Substituting the fit results for the form factors into the equation,
we calculate $I^l_K$ for the $K^0 \to \pi^- e^+ \nu_e$
and $K^0 \to \pi^- \mu^+ \nu_\mu$ processes and obtain each integral as,
\begin{eqnarray}
I^e_{K^0} &=& 0.15481(13)(^{\ +1}_{-11})(60)(3), \\
I^\mu_{K^0} &=& 0.10249(12)(^{\ +4}_{-16})(50)(3),
\end{eqnarray}
using $M_K = m_{K^0} = 0.497611$ GeV, $M_\pi = m_{\pi^-} = 0.13957061$ GeV, 
$m_e = 0.000511$ GeV, and $m_\mu = 0.10566$ GeV.
The result for each fitting form is presented in Table~\ref{tab:fit_chpt} and
Tables in Appendix~\ref{app:q2_fit_result}.
The central value, statistical and systematic errors are determined
in a similar way to the cases for the slope and curvature
as presented in the previous subsection,
and there is no systematic error coming from the choice of $Z_V$.
These results agree well with the experimental values in
the dispersive representation of the form factors,
$I^e_{K^0} = 0.15476(18)$ and $I^\mu_{K^0} = 0.10253(16)$,
in Ref.~\cite{Antonelli:2010yf}.

We also show the results for the unnormalized phase space integrals as,
\begin{eqnarray}
f_+(0) \sqrt{I^e_{K^0}} &=& 0.37783(62)(^{+54}_{-16})(171)(11)(9), \\
f_+(0) \sqrt{I^\mu_{K^0}} &=& 0.30742(49)(^{+35}_{-17})(139)(27)(9),
\end{eqnarray}
which will be used for evaluation of $|V_{us}|$ in the next section.
The errors are estimated in similar ways to the ones of $f_+(0)$.
Their numerical values for each fit form are summarized in 
Table~\ref{tab:fit_chpt} and
Tables in Appendix~\ref{app:q2_fit_result}.

\section{Result of $|V_{us}|$}
\label{sec:v_us}

Using our result of $f_+(0)$ in Sec.~\ref{sec:result_f0}
and the experimental value 
$|V_{us}|f_+(0) = 0.21654(41)$~\cite{Moulson:2017ive},
the result of $|V_{us}|$ in our study is given by
\begin{equation}
|V_{us}| = 0.22550(37)(^{+10}_{-34})(103)(43)(3)(43),
\label{eq:vus_f0}
\end{equation}
where the errors from the first to fifth inherit those
of $f_+(0)$ in Eq.~(\ref{eq:central_res_f00}).
The last error comes from the experimental one.
The result can be expressed as $|V_{us}| = 0.2255(13)(4)$,
where the first error is given by the combined error of
the five errors in our calculation.

\begin{figure}[ht!]
\makebox[.5\textwidth][r]{\includegraphics*[scale=.5]{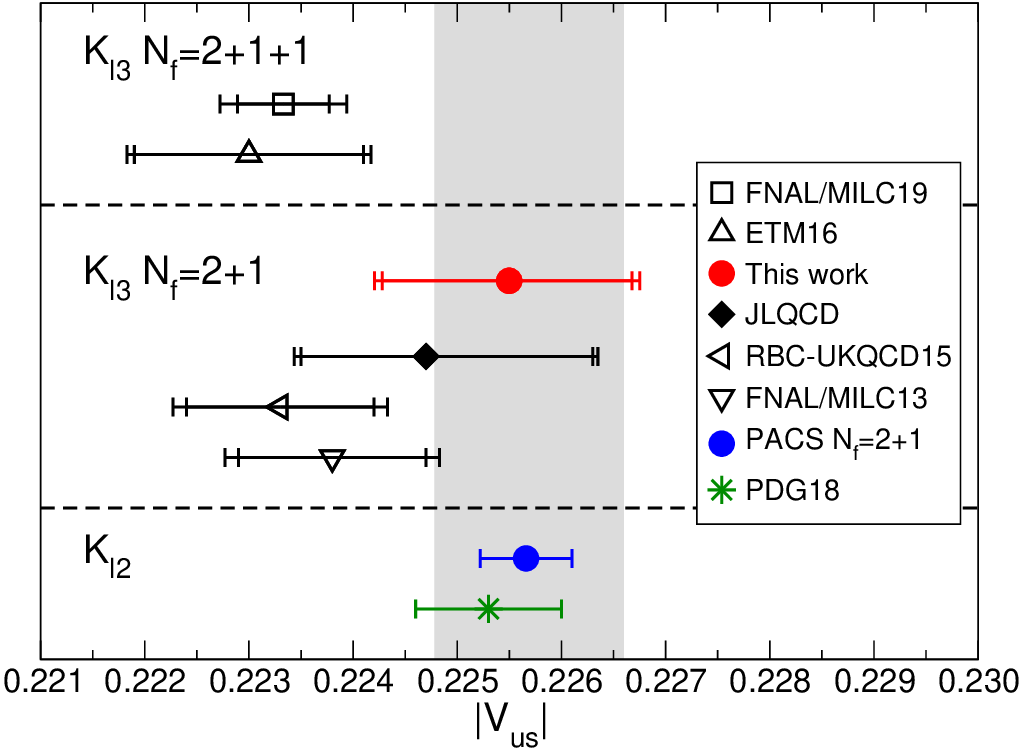}}
\caption{Comparison of $|V_{us}|$ with recent lattice QCD results obtained
from $K_{l3}$ form factors~\cite{Boyle:2013gsa,Boyle:2015hfa,Aoki:2017spo,Carrasco:2016kpy,Bazavov:2018kjg}
and the $K_{l2}$ decay using $F_K/F_\pi$ in 
our calculation~\cite{Ishikawa:2018jee} and PDG18~\cite{Tanabashi:2018oca}.
The inner and outer errors in the $K_{l3}$ calculations 
express the error of the lattice calculation and total error.
The total error is evaluated by adding the lattice QCD and experimental
errors in quadrature.
The unitarity value using $|V_{ud}|$ in PDG18~\cite{Tanabashi:2018oca} 
is presented by the shaded band.
The filled symbols represent the results at a finite lattice spacing.
}
\label{fig:comparison_vus}
\end{figure}

Figure~\ref{fig:comparison_vus} shows that our result 
is consistent with the value estimated by assuming 
the unitarity condition of the first row of the CKM matrix:
\begin{equation}
|V_{us}| = \sqrt{ 1 - |V_{ud}|^2 },
\end{equation}
where $|V_{ub}|$ is neglected due to $|V_{ub}| \ll |V_{ud}|$
and we use $|V_{ud}| = 0.97420(21)$~\cite{Marciano:2005ec,Tanabashi:2018oca}.
Furthermore, our result agrees with the results 
determined from the $K_{l2}$ decay process through
$|V_{us}|/|V_{ud}| \times F_K/F_\pi = 0.27599(38)$~\cite{Tanabashi:2018oca}.
In the figure we plot two data: one is obtained with the use of
the value of $F_K/F_\pi$ in PDG18~\cite{Tanabashi:2018oca}
and the other is from the result of $F_K/F_\pi$ calculated 
with the same configuration as in this work~\cite{Ishikawa:2018jee}.
These observations suggest that our result is consistent with 
the SM prediction within the error.
Using a new evaluation of $|V_{ud}| = 0.97370(14)$~\cite{Seng:2018yzq},
however, the value from the unitarity condition significantly changes as
$|V_{us}| = 0.2278(6)$, while the ones from the $K_{l2}$ decay do not move
within the error.
In this case, our result is smaller than the unitarity condition by
1.7 $\sigma$.
More recent evaluation of 
$|V_{ud}| = 0.97389(18)$~\cite{Czarnecki:2019mwq} leads to
the unitarity value of $|V_{us}| = 0.2270(8)$, which
is consistent with our result within 1.0 $\sigma$.

Figure~\ref{fig:comparison_vus} also presents the comparison of our result with 
the recent $N_f = 2+1$~\cite{Boyle:2013gsa,Boyle:2015hfa,Aoki:2017spo}
and $N_f = 2+1+1$~\cite{Carrasco:2016kpy,Bazavov:2018kjg} calculations.
Our result is reasonably consistent with all the results,
although it is 1.5 $\sigma$ larger than the recent result in 
Ref.~\cite{Bazavov:2018kjg} as for the result of $f_+(0)$ 
presented in Sec.~\ref{sec:result_f0}.
To understand the difference,
it is an important future work 
to reduce the uncertainties in our calculation.
We will remove our largest systematic error by
measuring the form factors at a finer lattice spacing
in the next calculation.

We also determine $|V_{us}|$ 
using the phase space integral calculated with our form factors,
$f_+(0) \sqrt{I^l_{K^0}}$ in Sec.~\ref{sec:phase_space_integral}.
The results for $l = e$ and $\mu$ are
\begin{eqnarray}
|V_{us}| &=& 
\left\{
\begin{array}{cc}
0.22524(37)(^{+10}_{-32})(103)(28)(5)(58) &  (l=e)\\ 
0.22558(36)(^{+13}_{-26})(103)(87)(6)(67) & (l=\mu)
\end{array}
\right.,
\end{eqnarray}
where we use $|V_{us}| f_+(0) \sqrt{I^e_{K^0}} = 0.08510(22)$
and $|V_{us}| f_+(0) \sqrt{I^\mu_{K^0}} = 0.06935(21)$,
which are evaluated from the experimental results and correction factors
in Ref.~\cite{Antonelli:2010yf}.
The meaning of the errors is the same as in the above $|V_{us}|$.
A weighted average of the two decay processes 
using the experimental error gives
\begin{equation}
|V_{us}| = 0.22539(37)(^{+11}_{-29})(103)(54)(6)(44).
\end{equation}
This value is well consistent with that in Eq.~(\ref{eq:vus_f0})
including the sizes for the uncertainties.
It is encouraging that the $K_{l3}$ form factors calculated
in the lattice QCD can be used for not only the determination
of $f_+(0)$, but also the evaluation of the phase space integral.

\section{Conclusion}
\label{sec:conclusions}

We have calculated the $K_{l3}$ form factors in
the $N_f = 2 + 1$ QCD at the physical point on the (10.9 fm)$^3$ volume
with the nonperturbatively $O(a)$-improved Wilson quark action and Iwasaki gauge action
at one lattice spacing corresponding to $a^{-1} = 2.3$ GeV.
Thanks to the large volume, we can access the form factors near $q^2 = 0$
without the twisted boundary technique.
For extraction of precise matrix elements for the $K_{l3}$ decay,
we have analyzed the corresponding 3-point functions avoiding 
excited state contributions,
such as the wrapping around effect of $\pi$ and the radial excited states
for $\pi$ and $K$.

To obtain the value of the form factors at $q^2 = 0$,
which is essential to evaluate $|V_{us}|$, 
we have interpolated the form factors to $q^2 = 0$
employing several fit forms. These interpolations 
also contribute to determination of the shape of the form factors as a function of
$q^2$.
The chiral extrapolation is not necessary in our analysis thanks to 
the calculation at the physical point.
The central value of the form factor at $q^2 = 0$ is determined 
with less than 0.2\% statistical error.
Since one of our data is very close to $q^2 = 0$,
the interpolations are fairly stable, and the systematic error 
from the interpolations is
as small as the statistical error.
The final result of $f_+(0)$ in this work is
\begin{equation}
f_+(0) = 0.9603(16)(^{+14}_{\ -4})(44)(19)(1),
\end{equation}
where the first error is statistical, and the second error
is the systematic one for the choice of the fit forms.
The third error is the largest systematic error in our result, which
comes from a finite lattice spacing effect 
estimated from the different determination of $Z_V$.
Another finite lattice spacing effect is estimated using higher order effect
of ChPT corresponding to the fourth error.
The isospin breaking effect in the fifth error is smaller compared 
to other ones.
Our result is reasonably consistent with the recent $N_f = 2 + 1$
and $N_f = 2 + 1 + 1$ QCD calculations.
The largest deviation from the recent results is 1.7 $\sigma$.
It is important to reduce the uncertainties in our calculation 
to understand the source of the deviation.
Thus, in the next calculation, we will measure the form factors
at a finer lattice spacing.

The slope and curvature for the form factors at $q^2 = 0$
are determined from the interpolations.
Their results are well consistent with the experimental values.
The interpolated result allows us to evaluate the phase space integral
and make a comparison with the experiment.
This evaluation can be regarded as comparison with the experimental
form factors in non-zero $q^2$ region.
Our values of the phase space integral agree with the experimental ones.

We have obtained $|V_{us}|$ from the result of $f_+(0)$ as,
\begin{equation}
|V_{us}| = 0.2255(13)(4),
\end{equation}
where the first error is the combined error in our calculation,
and the second comes from the experiment.
This result agrees with $|V_{us}|$ determined from the unitarity
condition of the CKM matrix and also from the $K_{l2}$ decay.
On the other hand, using a new evaluation of $|V_{ud}|$,
our result differs from the unitarity value by 1.7 $\sigma$.
To make the comparison with the SM predictions 
more stringent,
we would need to reduce the uncertainties in our calculation.
Thus, our next calculation with the finer lattice spacing 
is important also from this point of view.
Furthermore, nonperturbative evaluations of
the isosing breaking effect including the QED effect would be
important to search for BSM signal.

It is encouraging that another determination of $|V_{us}|$ using
the phase space integral evaluated with our form factors
completely agrees with the above conventional determination.
This suggests that the lattice calculation could contribute
to not only a precise determination of $f_+(0)$, but also 
the phase space integral.

\section*{Acknowledgments}
Numerical calculations in this work were performed on Oakforest-PACS
in Joint Center for Advanced High Performance Computing (JCAHPC)
under Multidisciplinary Cooperative Research Program of Center for Computational Sciences, University of Tsukuba.
A part of the calculation employed OpenQCD system\footnote{http://luscher.web.cern.ch/luscher/openQCD/}.
This work was supported in part by Grants-in-Aid 
for Scientific Research from the Ministry of Education, Culture, Sports, 
Science and Technology (Nos. 16H06002, 18K03638, 19H01892).

\appendix

\section{Functions of NLO ChPT formulae}
\label{app:chpt_formulae}

The functions $K_+(q^2)$ and $K_0(q^2)$ in Eqs.~(\ref{eq:NLO_chpt_f+}) and
(\ref{eq:NLO_chpt_f0}) are summarized in this appendix,
which appears in the NLO SU(3) ChPT 
formulae~\cite{Gasser:1984ux,Gasser:1984gg}.

The function $K_+(q^2)$ of $f_+(q^2)$ in Eq.~(\ref{eq:NLO_chpt_f+})
is given by
\begin{equation}
K_+(q^2) = \frac{3}{32\pi^2 F_0^2}
( H_{\pi K}(t) + H_{K\eta}(t) ),
\end{equation}
with $t = -q^2$ and
\begin{eqnarray}
H_{ab}(t) &=& 
\frac{1}{12}\left(t - 2 \Sigma_{ab} + \frac{\Delta_{ab}^2}{t}\right)J_{ab}(t)
- \frac{1}{3}J^\prime_{ab}
- \frac{t}{6}k_{ab} + \frac{t}{9},
\end{eqnarray}
where
\begin{eqnarray}
J_{ab}(t) &=& 2 - \left(\frac{\Delta_{ab}}{t}-\frac{\Sigma_{ab}}{\Delta_{ab}}\right)
\ln\frac{m_a^2}{m_b^2} - \frac{\nu}{t}\ln 
\frac{(t+\nu)^2 - \Delta_{ab}^2}{(t-\nu)^2 - \Delta_{ab}^2},\\
J^\prime_{ab} &=& \Sigma_{ab} + \frac{2 m_a^2 m_b^2}{\Delta_{ab}}
\ln \frac{m_b^2}{m_a^2},\\
k_{ab} &=& \frac{\mu_a - \mu_b}{\Delta_{ab}},\\
\Sigma_{ab} &=& m_a^2 + m_b^2,\\
\Delta_{ab} &=& m_a^2 - m_b^2,
\end{eqnarray}
with
\begin{eqnarray}
\nu^2 &=& \Delta_{ab}^2 - 2 \Sigma_{ab} t + t^2,\\
\mu_a &=& m_a^2 \ln ( m_a^2 / \mu^2 ) .
\end{eqnarray}
$K_0(q^2)$ of $f_0(q^2)$ in Eq.~(\ref{eq:NLO_chpt_f0}) is defined by
\begin{eqnarray}
K_0(q^2) &=& \frac{1}{32\pi^2 F_0^2}
\left\{ 
\frac{1}{4}\left( 5 t - 2 \Sigma_{\pi K} - \frac{3 \Delta_{\pi K}^2}{t}\right)
J_{\pi K}(t) 
+ \frac{1}{12}\left( 3 t - 2 \Sigma_{\pi K} - \frac{\Delta_{\pi K}^2}{t}\right)
J_{K \eta}(t)\right. \nonumber\\
&&
\left.
- \frac{t}{2 \Delta_{\pi K}} \left( 5 \mu_\pi - 2 \mu_K - 3 \mu_\eta \right)
\right\} .
\end{eqnarray}
We adopt $m_\eta^2 = ( 4 m_K^2 - m_\pi^2 ) / 3$ and $\mu = 770$ MeV.
The two functions give the same value at $q^2 = 0$, $K_+(0) = K_0(0)$,
in this choice of $m_\eta^2$.

\section{Results of $q^2$ interpolation}
\label{app:q2_fit_result}

In this appendix several fit results for the interpolations of the form factors
are summarized.
In addition to the fit forms based on the NLO SU(3) ChPT formulae 
in Eqs.~(\ref{eq:NLO_chpt_f+}) and (\ref{eq:NLO_chpt_f0}),
we also employ several fit forms for the interpolation, 
such as mono-pole functions,
\begin{equation}
f_+(q^2) = \frac{f_+(0)}{q^2 + M_+^2} \ \ {\rm and}\ \ 
f_0(q^2) = \frac{f_+(0)}{q^2 + M_0^2} ,
\label{eq:monopole_q2}
\end{equation}
simple quadratic functions of $q^2$, 
\begin{equation}
f_+(q^2) = f_+(0) + c_1^+ q^2 + c_2^+ q^4
\ \ {\rm and}\ \ 
f_0(q^2) = f_+(0) + c_1^0 q^2 + c_2^0 q^4 ,
\label{eq:quadratic_q2}
\end{equation}
and variations of the $z$-parameter expansion~\cite{Bourrely:2008za},
\begin{equation}
f_+(q^2) = f_+(0) + \sum_{i=1}^{N_z} c_i^+ z^i(q^2)
\ \ {\rm and}\ \ 
f_0(q^2) = f_+(0) + \sum_{i=1}^{N_z} c_i^0 z^i(q^2)
\label{eq:zexpansion_q2}
\end{equation}
where $N_z = 1$ or 2 and
\begin{equation}
z(q^2) = \frac{\sqrt{(m_K+m_\pi)^2+q^2} - (m_K+m_\pi)}
{\sqrt{(m_K+m_\pi)^2+q^2} + (m_K+m_\pi)} .
\end{equation}
Our choice of $z(q^2)$ corresponds to the one with $t_0 = 0$ in the general representation of $z(q^2)$~\cite{Bourrely:2008za}.
The parameters $f_+(0)$, $c_i^{+,0}$, and $M_{+,0}$ are 
fit parameters.
These fit results are summarized in 
Tables~\ref{tab:fit_monopole}--\ref{tab:fit_zexpansion}.

\begin{table}[ht!]
\caption{Fit results of $K_{l3}$ form factors using monopole fit forms
in Eq.~(\ref{eq:monopole_q2})
together with the values of the uncorrelated $\chi^2/$dof.
A1 and A2 data sets are explained in Sec.~\ref{sec:tsep}.
We also list the results for slope, curvature, and 
phase space integral.}
\label{tab:fit_monopole}
\begin{tabular}{ccc}\hline\hline
& A1 & A2\\\hline
$f_+(0)$ & 0.9599(16) & 0.9612(24) \\
$M_+^2$ [GeV$^{2}$] & 1.311(11) & 1.309(16) \\
$M_0^2$ [GeV$^{2}$] & 0.7210(92) & 0.714(16) \\
$\chi^2/$dof & 0.16 & 0.24 \\
$\lambda_+^\prime$ [$10^{-2}$] & 2.554(21) & 2.550(30)\\
$\lambda_0^\prime$ [$10^{-2}$] & 1.405(18) & 1.390(32)\\
$\lambda_+^{\prime\prime}$ [$10^{-3}$] & 1.359(23) & 1.353(34) \\
$\lambda_0^{\prime\prime}$ [$10^{-3}$] & 0.411(11) & 0.402(19) \\
$I^e_{K^0}$ & 0.15475(13) & 0.15472(19) \\
$I^\mu_{K^0}$ & 0.10251(12) & 0.10246(18)\\
$f_+(0)\sqrt{I^e_{K^0}}$ & 0.37759(61) & 0.37810(87) \\
$f_+(0)\sqrt{I^\mu_{K^0}}$ & 0.30732(48) & 0.30768(64)
\\\hline\hline
\end{tabular}
\end{table}

\begin{table}[ht!]
\caption{Same as Table~\ref{tab:fit_monopole}, but for
quadratic fit forms in Eq.~(\ref{eq:quadratic_q2}).}
\label{tab:fit_polynomial}
\begin{tabular}{ccc}\hline\hline
& A1 & A2\\\hline
$f_+(0)$ & 0.9600(16) & 0.9615(24) \\
$c_1^+$ [GeV$^{-2}$] & $-$1.283(18) & $-$1.296(35) \\
$c_2^+$ [GeV$^{-4}$] & 1.45(17)     & 1.31(36) \\
$c_1^0$ [GeV$^{-2}$] & $-$0.677(18) & $-$0.664(33) \\
$c_2^0$ [GeV$^{-4}$] & 0.68(12)     & 0.71(19) \\
$\chi^2/$dof & 0.05 & 0.18 \\
$\lambda_+^\prime$ [$10^{-2}$] & 2.620(37) & 2.642(70)\\
$\lambda_0^\prime$ [$10^{-2}$] & 1.383(37) & 1.353(69)\\
$\lambda_+^{\prime\prime}$ [$10^{-3}$] & 1.16(14) & 1.05(29) \\
$\lambda_0^{\prime\prime}$ [$10^{-3}$] & 0.543(94) & 0.57(15) \\
$I^e_{K^0}$ & 0.15480(12) & 0.15483(22) \\
$I^\mu_{K^0}$ & 0.10249(12) & 0.10245(21)\\
$f_+(0)\sqrt{I^e_{K^0}}$ & 0.37769(61) & 0.37835(93) \\
$f_+(0)\sqrt{I^\mu_{K^0}}$ & 0.30732(49) & 0.30776(68)
\\\hline\hline
\end{tabular}
\end{table}

\begin{table}[ht!]
\caption{Same as Table~\ref{tab:fit_monopole}, but for
fits using variations of the $z$-parameter expansion
in Eq.~(\ref{eq:zexpansion_q2}) with $N_z = 1$ and 2.
}
\label{tab:fit_zexpansion}
\begin{tabular}{cccccc}\hline\hline
& \multicolumn{2}{c}{A1} & \multicolumn{2}{c}{A2}\\
$N_z$ & 1 & 2 && 1 & 2 \\\hline
$f_+(0)$ & 0.9598(16)     & 0.9601(16)   && 0.9615(23) & 0.9617(24) \\
$c_1^+$ & $-$2.017(18) & $-$2.045(29) && $-$2.013(26) & $-$2.073(60) \\
$c_2^+$ & $\cdots$     & $-$0.75(40)  && $\cdots$     & $-$1.21(88)  \\
$c_1^0$ & $-$1.031(14) & $-$1.084(27) && $-$1.012(23) & $-$1.066(47) \\
$c_2^0$ & $\cdots$     & $-$0.70(23)  && $\cdots$     & $-$0.66(34)  \\
$\chi^2/$dof & 0.34 & 0.04 && 0.35 & 0.19 \\
$\lambda_+^\prime$ [$10^{-2}$] & 2.576(23) & 2.612(38) && 2.566(34) & 2.643(74) \\
$\lambda_0^\prime$ [$10^{-2}$] & 1.318(18) & 1.384(35) && 1.290(32) & 1.359(62)\\
$\lambda_+^{\prime\prime}$ [$10^{-3}$] & 1.263(12) & 1.05(11) && 1.258(17) & 0.92(24) \\
$\lambda_0^{\prime\prime}$ [$10^{-3}$] & 0.6461(91) & 0.458(56) && 0.633(16) & 0.460(81)\\
$I^e_{K^0}$ & 0.15476(13) & 0.15475(13) && 0.15470(19) & 0.15480(23) \\
$I^\mu_{K^0}$ & 0.10243(12) & 0.10246(12) && 0.10233(18) & 0.10242(22) \\
$f_+(0)\sqrt{I^e_{K^0}}$ & 0.37758(61) & 0.37768(61) && 0.37817(87) & 0.37836(95) \\
$f_+(0)\sqrt{I^\mu_{K^0}}$ & 0.30717(48) & 0.30709(49) && 0.30757(64) & 0.30776(69)
\\\hline\hline
\end{tabular}
\end{table}

\bibliographystyle{apsrev4-1}
\bibliography{reference}

\end{document}